\documentclass[fleqn,usenatbib]{mnras}

\usepackage{newtxtext,newtxmath}
\usepackage[T1]{fontenc}

\DeclareRobustCommand{\VAN}[3]{#2}
\let\VANthebibliography\thebibliography
\def\thebibliography{\DeclareRobustCommand{\VAN}[3]{##3}\VANthebibliography}

\usepackage{graphicx}	% Including figure files
\usepackage{amsmath}	% Advanced maths commands
\usepackage[normalem]{ulem}

\newcommand{\ramses}{{\rm \small RAMSES}\ }

\newcommand{\ramsesrt}{{\rm \small RAMSES-RT}\ }

\newcommand{\rascas}{{\rm \small RASCAS}\ }

\newcommand{\Lya}{${\rm Ly\alpha}$}
\newcommand{\Ha}{${\rm H\alpha}$}

\newcommand{\Msolyr}{{\rm\, M_\odot\,yr^{-1}}}
\newcommand{\Msol}{{\rm\, M_\odot}}

\newcommand{\HI}{$\rm H\small{I}$}
\newcommand{\HII}{$\rm H\small{II}$}

\newcommand{\kmsec}{${\,\rm {km\,s^{-1}} }$}

\title[]{Radiation hydrodynamic simulation of the Haro 11 galaxy: the escape of LyC and Ly$\alpha$ in a dwarf galaxy merger}
\author[T. Ejdetjärn et al.]{
Timmy Ejdetjärn$^{1}$\thanks{E-mail:timmy.ejdetjarn@gmail.com}, Göran Östlin$^{1}$, Joakim Rosdahl$^{2}$, Jérémy Blaizot$^{2}$, Oscar Agertz$^{3}$
\\\\
$^{1}$Oskar Klein Centre, Department of Astronomy, Stockholm University, 106 91 Stockholm, Sweden\\
$^{2}$Centre de Recherche Astrophysique de Lyon UMR5574, Univ Lyon, Univ Lyon1, Ens de Lyon, CNRS, F-69230 Saint-Genis-Laval, France\\
$^{3}$Lund Observatory, Division of Astrophysics, Department of Physics, Lund University, Box 43, SE-221 00 Lund, Sweden\\
}

\date{Accepted XXX. Received YYY; in original form ZZZ}

\pubyear{2026}

\begin{document}
\label{firstpage}
\pagerange{\pageref{firstpage}--\pageref{lastpage}}
\maketitle

\begin{abstract}
% Local starburst galaxies are thought to be analogues of high-redshift galaxies that contributed to the ionisation of the universe. 
The Haro 11 galaxy merger is the closest known Lyman Continuum (LyC) leaker and a strong Lyman-$\alpha$ (Ly$\alpha$) emitter, making it an important analogue of the high-$z$ galaxies that reionised the early Universe. To investigate how Haro 11's properties arise, we perform a radiation hydrodynamics simulation of the merger, and create mock observations of LyC, Ly$\alpha$, and H$\alpha$, from which we compute their luminosities ($L$) and escape fractions ($f_{\rm esc}$). We track these quantities along multiple sightlines as the two progenitor galaxies merge, from the first interaction until the system resembles present-day Haro 11. We find that $L$ and $f_{\rm esc}$ vary by 1-2 orders of magnitude for LyC due to sightline variations. At the two pericentre passages, the total $f_{\rm esc}^{\rm LyC}$ increases by roughly an order of magnitude. Conversely, $f_{\rm esc}^{\rm Ly\alpha}$ shows a moderate increase at the pericentre passages, which affects the inference of LyC properties from Ly$\alpha$. We attribute this to a displacement of the LyC-emitting stars relative to the \Lya-emitting gas, combined with an increased density from gas compression. Furthermore, $f_{\rm esc}^{\rm LyC}$ is boosted during star formation bursts, likely due to stellar feedback. As direct comparison with Haro 11, the simulation qualitatively matches its morphology and luminosities. We find that among the dense stellar knots, knot C is the main contributor to both intrinsic and escaping LyC emission. Additionally, the Ly$\alpha$ spectra displays distinct features found in observations, implying similar gas conditions are present.
% Compared to observations, our simulation yields a qualitative match to the morphology and luminosities of Haro 11. 
\end{abstract}


%%%%%%%%%%%%%%%%%%%%%%%%%%%%%%%%%%%%%%%%%%%%%%%%%%
%%%%%%%%%%%%%%%%% BODY OF PAPER %%%%%%%%%%%%%%%%%%
%%%%%%%%%%%%%%%%%%%%%%%%%%%%%%%%%%%%%%%%%%%%%%%%%%

%%%%%%%%%%%%%%%%%%%%%%%%%%%%%%%%%%%%%%%%%%%%%%%%
\section{Introduction}

%%%%%%%%%%%%%%%%%%%%%%%%%%%%%%%%%%%%%%%%%%%%%%%%%%%%%%
%%%%%%%%%%%%%%%%%%%%%%%%%%%%%%%%%%%%%%%%%%%%%%%%%%%%%%
Understanding the processes that governed the reionisation of the Universe is one of the central challenges in modern astrophysics. The transition of the intergalactic medium (IGM) from a neutral to ionised state, by redshift $z\sim6$ \citep[e.g.][]{McGreer+15, Bosman+22}, is believed to be primarily driven by early starburst galaxies \citep[][]{Robertson+15, Matsuoka+18, Matthee+22, Choustikov+24}. In particular, low-mass galaxies with ionising escape fracions $\sim10-20\,\%$ have been suggested to be crucial in this process \citep[][]{Anderson+17, Finkelstein+19, Lewis+20, Simmonds+24, Atek+24}. However, the constraints at high redshift are exclusively theoretical: observational studies of ionising radiation in galaxies at $z>4$ are hampered by the opaque neutral gas filling the intergalactic medium \citep[][]{Madau1995,InoueIwata2008,Inoue+14}. An alternative observational avenue is local low-mass starburst galaxies, which have been suggested to be analogue to those driving reionisation, due to their high star formation rates (SFR), low metallicities, and compact morphologies \citep[e.g.][]{Ostlin+01, Schaerer+16, Gao+22}.

Until recently, only a few local analogues were directly observed to leak ionising Lyman Continuum (LyC) radiation, e.g. Mrk54 \citep[][]{Deharveng+01}, Haro 11 \citep[][]{Bergvall+06}, and Tol 1247 \citep[][]{Leitet+13}; see also \citet[][]{Izotov+18} and references therein. The Low-Redshift Lyman Continuum Survey \citep[LzLCS;][e.g.]{Wang+21, Saldana-Lopez+22, Flury+22a, Flury+22b, Omkar+24} more than doubled the number of previously known local ($z<0.4$) Lyman Continuum emitters (LCEs) with $f_{\rm esc}^{\rm LyC}> 3\,\%$. These are primarily compact galaxies with high SFR column densities, but exhibit diverse morphologies and varying escape fractions. This recent boost in local LCEs have made comparison with high-$z$ observations and simulations more feasible.

The Haro 11 galaxy is the closest observed \citep[$z\sim0.02$, or $d\sim88.5$\,Mpc;][]{Sirressi+22} LCE \citep[][]{Bergvall+06, Rivera-Thorsen+17a, Ostlin+21, Komarova+24} with a mean $f_{\rm esc}^{\rm LyC}\sim3-4$\,\% \citep[][]{Leitet+11, Komarova+24}. Additionally, it is a blue, compact (dwarf) galaxy (BCD/BCG), currently undergoing a starburst with an SFR of $\sim30\Msolyr$ \citep[][]{Hayes+07, Adamo+10, Madden+13, MacHattie+14, Gao+22} likely the result of an ongoing merger (see \citealt{Ostlin+15} and \citealt{Ejdetjarn+25} for detailed overviews). All of these conditions make Haro 11 a strong analogue for high-$z$ LCE galaxies.

Additionally, it was recently discovered that Haro 11 has an \HI{} tidal tail \citep[][]{LeReste+24}, which is a common feature of galaxy mergers, and a similar disturbed gas morphology has been found in several LCEs \citep[e.g.][]{Bergvall+13, Leitet+13, Leitherer+16, LeReste+25c}. This indicates that processes which offset neutral gas from the LyC source, such as gas stripping and disturbed morphology, might play a role in facilitating LyC escape. Nevertheless, gas clearing and ionisation from stellar feedback has been found to be sufficient to explain the LyC emission in many cases \citep[e.g.][]{Trebitsch+17, Kimm+17, Flury+24}.

%, in particular for Haro 11 which shows pockets of reduce neutral gas and \citep[][]{LeReste+24}. 
% might indicate that neutral gas stripping plays a role in forming the conditions for LyC leaking

Compared to the local Universe, mergers were more frequent in the past, as indicated by both observations \citep[e.g.][]{Duncan+19} and simulations \citep[e.g.][]{Rodriguez-Gomez+15}. Major mergers are well-known mechanisms that tidally ejects gas and stars, and create compact irregular morphologies, thus increasing their SFRs and LyC production. This highlights the relevance of mergers for facilitating the escape of radiation during the epoch of reionisation, but the exact impact of mergers has not yet been established. Recently, \citet[][]{Mascia+25} used JWST data from various surveys to evaluate the correlation between morphology and escape fraction, for galaxies at $5\leq z\leq7$ (using correlations from LzLCS to derive LyC production and escape). They reported a mean $f_{\rm esc}^{\rm LyC}\sim4\,\%$ for their sample and found no correlation between high escape fractions and disturbed morphologies; instead they attributed boosts in $f_{\rm esc}^{\rm LyC}$ to SFR bursts and compactness. Conversely, \citet[][]{Zhu+24} found that majority of LCE galaxies in the GOODS-S field at $z\gtrsim3$ show signs of recent mergers and, similarly, \citet[][]{LeReste+25c} demonstrated that local galaxies part of the LaCOS survey show signs of recent mergers. Additionally, post-processing of the TNG50 large-scale cosmological simulations by \citet{KostyukCiardi2024} indicated that LyC escape is typically boosted by a factor of a few following a merger, due to an increase in SFR and displacement of neutral gas.

% Cosmological simulations have offered a great way to study the large-scale effects of re-ionisation, but understanding the small scale effects that facilitate the escape remains elusive. Zoom simulation have found that ...

Due to the limitations of observing ionising radiation in high-$z$ galaxies, observers often have to rely on indirect measurements of the LyC properties, such as strong emission lines (e.g. \Lya{}, {\rm [O\,{\small III}],\ [O\,{\small II}]). In particular, observational studies of low-redshift galaxies have shown that strong \Lya{} emission indicates LyC leakage \citep[][]{Verhamme+17, Steidel+18, Vanzella+18, Izotov+21, Saldana-Lopez+23, Solhaug+25}. Recently, the LaCOS survey \citep[][]{LeReste+25b, Saldana-Lopez+25, LeReste+25c} found various correlations between LyC and \Lya{} properties in local galaxies part of the LzLCS. However, several galaxies with prominent \Lya\ emission have been found to be weak LyC leakers \citep[e.g.][]{Naidu+22, Solhaug+25, Citro+25}.

% For the LzLCS sample, \citet[][]{Flury+24} demonstrated that sightlines with dense clumps 
% angle variation of LyC in the local Universe and mechanical feedback from supernovae $\lesssim 10$\,Myr are important to drive the anisotropy for LyC leakage. 

%%% FLury+22: This concentration at low fesc Lyα suggests a high H I column along the line of sight in many of the LzLCS galaxies (e.g., Verhamme et al. 2015).
%For example, \citet[][]{Maji+22} found that brighter \Lya\ emitters are the more dominant contributors of LyC escape into the IGM. 
This relation between \Lya\ and LyC is predicted by theoretical transfer equations \citep[e.g.][]{Verhamme+15} and has been found in large-scale simulations. \citet[][]{Choustikov+24} used the SPHINX$^{20}$ cosmological simulations to study the escape of LyC and \Lya\ in high redshift ($z\sim 5$) galaxies. The authors showed that $f_{\rm esc}^{\rm Ly\alpha}$ is a good diagnostic for LyC leakage, as it selects galaxies with low neutral gas densities, and that a high $f_{\rm esc}^{\rm LyC}$ is correlated with a less extended \Lya\ distribution. This behaviour was recently confirmed by the LaCOS survey \citep[][]{Saldana-Lopez+25} as well as observations of extended Mg{\small II} halos and LyC escape \citep[][]{Leclercq+24}. Also using the SPHINX suite, \citet[][]{Maji+22} found that brighter \Lya\ emitters are the more dominant contributors of LyC escape into the IGM, but with a lot of scatter from galaxy to galaxy. Additionally, LyC escape is highly dependent on the column density and clump geometry \citep[e.g.][]{Mauerhofer+21, Flury+24}, which can vary widely between sightlines and over time. \citet[][]{Rosdahl+22} demonstrated that in the {\small SPHINX} suite the LyC varies highly over time-scales of only a few Myr and is strongly regulated by stellar feedback to produce escape channels.

% For the Lyman Alpha Reference Sample (LARS), \citet[][]{LeReste+25} reported little to no correlation between the brightness in \Lya\ and global \HI\ properties. 
% evaluate the correlation between \Lya\ and \HI properties in galaxies, and 

As discussed, few numerical works in literature studying \Lya{} or LyC escape have evaluated the effects of galaxy mergers. All of these papers have used cosmological simulations, which can track the overall effects of mergers over cosmic time. However, this approach presents difficulties in isolating the effects of a single merger from other cosmological influences, e.g. the circumgalactic medium, gas accretion, or other (simultaneous) gravitational interactions. In this paper, we perform a high-resolution simulation of a galaxy merger between two discs in an isolated space, which offers a well-controlled environment where the direct effect of the merger can be easily disentangled and understood.
% only followed the overall impact of mergers which may vary widely between specific mergers

In \citet[][hereafter E25]{Ejdetjarn+25}, we employed hydrodynamical simulations to demonstrate that the Haro 11 galaxy can be described as a merger between two disc galaxies, by reproducing a wide range of different observed features of the galaxy. By matching our simulation with observations, it allows for further analysis of other features and their possible origin. In this paper, we elaborate on the results of the previous paper by studying the production and escape of ionising radiation from our galaxy during the merger. The simulations presented have been run at a higher spatial resolution and include radiative transfer on-the-fly, which allows us to trace the propagation of ionising radiation and make mock observations using the post-processing tool \rascas \citep[][]{Michel-Dansac+20}.

 % Furthermore, isolated galaxies require less computational time, allowing for high resolution simulations to be run.

% In this paper, we model the merger of two disc galaxies to reproduce the general morphology and properties of the Haro 11 galaxy as observed today, and discuss a possible origin of its formation. We perform a scheme of low-resolution hydrodynamical and $N$-body simulations of the interaction, varying the galaxy and orbital parameters until the merging galaxies' morphology resembles the morphology of Haro 11 today. Using various observational studies of the gaseous and stellar components, e.g. star formation rate and masses, to constrain the initial conditions our model. We decide on a fiducial simulation that best represents Haro 11 and run it at a higher resolution, at 9 pc. We present the results of this fiducial simulation and compare the successes and shortcomings of our models and discuss its applications.

The paper is structured as follows. In Section~\ref{sec:simulations}, we describe the simulation setup, subgrid models, post-processing, and our method for producing mock observations and calculating values. In Section~\ref{sec:results}, we visualise the gaseous and stellar properties of the simulated galaxy, as well as emission maps of LyC, \Lya, and \Ha{} at present-day (i.e. the snapshot most resembling Haro 11 today). Additionally, we present the intrinsic and escaping luminosities as a function of time. In Section~\ref{sec:discussion}, we present the mock observations of the luminosities and escape fractions over time as observed from many different sightlines. Additionally, by constructing resolved maps of the \Lya\ line profile and escape fractions, we are able to comment on the properties of the emissions from the different locations (or 'knots') in the galaxy. At the end of this section we discuss possible caveats to our results. Finally, we conclude our results in Section~\ref{sec:conclusions}, with an outlook for the future.

%Starburst galaxies, characterized by their high rates of star formation, play a critical role in understanding the evolution of the universe, particularly during the epoch of reionisation. This period, occurring roughly between 500 million and 1 billion years after the Big Bang, marked the universe’s transition from a neutral to an ionized state. High-energy photons from newly forming stars in these galaxies are thought to have contributed significantly to reionising the intergalactic medium \citep[][]{Robertson+15}. Investigating galaxies that exhibit intense star formation today allows astronomers to better model the processes that may have driven this early phase of cosmic history.

%Haro 11 is a nearby starburst galaxy located at a redshift of $z \sim 0.02$, and has been a crucial laboratory for understanding star formation and ionizing photon leakage, both of which are integral to the reionisation process. Haro 11 is especially notable due to its similarities to the young, compact, and high-redshift galaxies thought to have been responsible for reionising the universe \citep[][]{Ostlin+15}. Its high ultraviolet (UV) luminosity and the detection of ionizing radiation escaping from the galaxy make it a key object of study. Furthermore, Haro 11 exhibits characteristics of both starburst and Lyman-continuum-leaking galaxies, which strengthens its relevance to investigations into how early galaxies may have driven the reionisation process \citep[][]{Pardy+16}.

\begin{figure*}
    \includegraphics[width=0.51\textwidth]{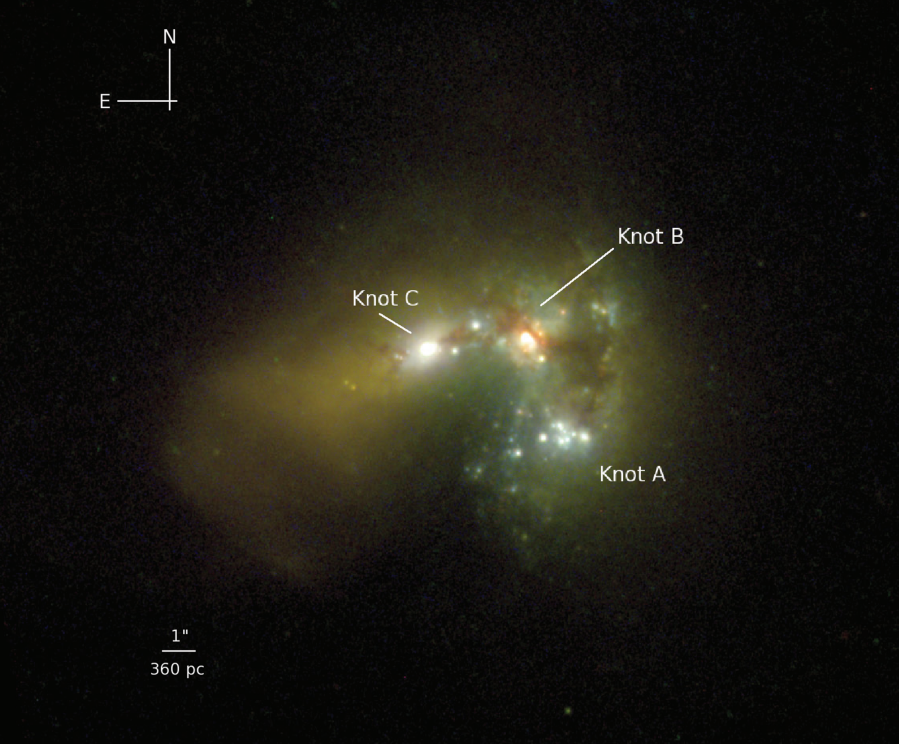}
    \includegraphics[width=0.42\textwidth]{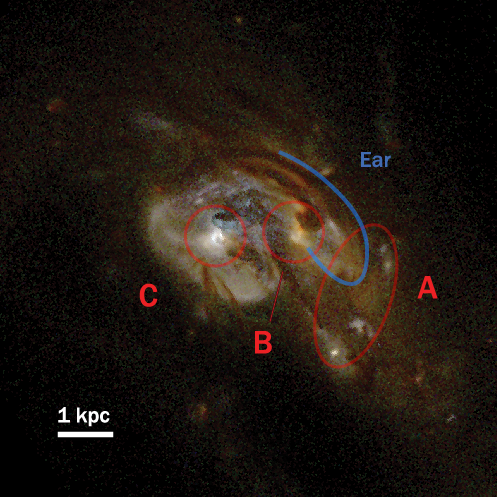}
    \caption{\emph{Left:} Composite HST image of Haro 11 observed in 220W, 435W, and 814W \citep[image from][]{Adamo+10}. \emph{Right:} Mock observations of our simulated Haro 11, using \rascas to apply the same HST filters (see Section~\ref{sec:mock_method} for details on how this image was produced).}
    \label{fig:Haro11_cont}
\end{figure*}

%%%%%%%%%%%%%%%%%%%%%%%%%%%%%%%%%%%%%%%%%%%%%%%%%%
\section{Numerical methods} \label{sec:simulations}
The numerical work presented here uses the radiation hydrodynamics code (RHD) \ramsesrt \citep[][]{Rosdahl+13,RosdahlTeyssier15}, built on the base \ramses code \citep[][]{Teyssier2002}. The gaseous component is tracked on a non-uniform grid using Adaptive Mesh Refinement (AMR), and its dynamics follow the conservative Euler equations. The gaseous fluid is assumed to be an ideal mono-atomic gas with adiabatic index $\gamma = 5/3$, and experiences hydrodynamical effects, radiation transfer, non-equilibrium cooling/heating, and gravitational effects from gas, stars, and dark matter particles. The hydrodynamic equations are solved with an HLLC Riemann solver \citep[][]{ToroSpruceSpeares94} and using a 2nd order Godunov scheme.

In the coming sections, we briefly outline the sub-grid method used for star formation and stellar feedback, but direct the reader to \citet[][]{Agertz+20, Agertz+21} for a comprehensive description of the physical recipes \citep[see also][for details about the implementation]{Agertz+13}. Furthermore, we summarise the RT setup employed in our simulation, with full details provided in \citet[][see also \citealt{RosdahlTeyssier15}]{Rosdahl+13}{}{}, and give an overview of the post-processing with \rascas \citep[see][for specifics]{Michel-Dansac+20}. Finally, we briefly describe the setup of the two merging galaxies, but refer to E25 for a full account of how the initial properties and orbital parameters were determined.

% cloud-in-cell interpolation \citep[][]{GuilletTeyssier11}{}{}. 

%%%%%%%%%%%%%%%%%%
\subsection{Star formation and feedback physics}
Star formation is treated as stochastic events, and the amount of stars formed is sampled from a discrete Poisson distribution. The initial mass of a star particle is $1000\Msol$, and represents a single stellar population with a \citet{Chabrier03} initial mass function (IMF). Star formation is set by the Schmidt law \citep[][]{Schmidt59, Kennicutt98} as 
\begin{align}
    \dot\rho_* = \epsilon_{\rm ff} \frac{\rho_{\rm g}}{t_{\rm ff}}\ {\rm for}\ \rho_{\rm g} > \rho_*,
\end{align}
where  $\rho_* = 100\,m_{\rm H}\,{\rm cm^{-3}}$ is the density threshold for star formation, $\rho_{\rm g}$ is the local gas density, $t_{\rm ff} = \sqrt{3\pi/32 G \rho}$ is the free-fall time for a spherical cloud (with gas \& stellar density $\rho$), and $\epsilon_{\rm ff}$ is the star formation efficiency per free-fall time. For our simulations, we adopt $\epsilon_{\rm ff} = 10\%$, which galaxy simulations \citep[see e.g.][]{AgertzKravtsov16,Grisdale2018,Grisdale2019} have shown, given strong feedback, to reproduce the $\epsilon_{\rm ff} \sim 1\%$ observed at kpc-scale average in galaxies \citep[e.g.][]{Bigiel+08} and found in molecular cloud observations \citep[e.g.][]{KrumholzTan07}.

The stellar feedback effects considered in this work are stellar winds, supernovae (Type Ia \& II), and radiation (through the RT module). For winds and supernovae, we track the injection of thermal and kinetic energy, momentum, and metals, from stellar particles back into the interstellar medium (ISM) through a subgrid recipe \citep[see][]{Agertz+13,AgertzKravtsov15}; based on their age, mass, and metallicity. To fully capture supernovae momentum, the explosion is modelled by injecting thermal energy into the host cell if the supernova is resolved, and otherwise the momentum is directly injected into the surrounding gas cells. A supernova is considered resolved if the host cell is larger than at least six cooling lengths \citep[following][see also \citealt{Agertz+21} for implementation details of our simulation]{KimOstriker15}. Metallicity is traced as a passive hydro scalars of oxygen and iron, which are then combined into a total metallicity \citep[following relative Solar abundances;][]{Asplund2009} and used for cooling, heating, and RT calculations. Radiation feedback is handled self-consistently by the RT module, which we will detail next.

%%%%%%%%%%%%%%%%%%%%%%%%%%%
\subsection{Radiative transfer}\label{sec:RT}
\ramsesrt uses a moment-based radiative transfer method, which allows for radiation to be diffused on the AMR grid following a set of conservation laws. Energy and momentum are transferred through both single and multiple scattering processes \citep[][]{Rosdahl+15}. The code applies a M1 closure approximation to connect the radiation pressure through the local energy and flux, using the Eddington tensor \citep[][]{Levermore84}. The flux between grid cells is calculated with the Global Lax-Friedrichs intercell flux \citep[see][]{Rosdahl+13}{}{}. We adopt a reduced speed of light $\bar c = 0.005\,c$, in order to reduce the computational times due to Courant-Friedrichs-Lewy timestepping condition\footnote{ Brief testing showed no difference in the galaxy ISM when increasing this to 0.02 or 1.00.}. The radiation energy emitted by a stellar particle is described by the spectral energy distribution emitted by a single-age stellar population, assuming a Chabrier IMF taken from \citet[][]{BruzualCharlot03}.

The radiation is tracked as wavepackets, which are grouped into photon groups of specific frequency ranges, defined to represent important radiative interaction within the ISM. We adopt four photon groups that cover: infrared radiation, ionising radiation of {\rm H{\small I}}, {\rm He{\small I}}, and {\rm He{\small II}} \citep[see][for details]{Rosdahl+15}. Additionally, the number fraction of the ion species ($x_{\rm H \small II }$,  $x_{\rm He \small I }$, $x_{\rm He \small II }$) are tracked as passive scalars on the AMR mesh and are solved through non-equilibrium chemistry and radiative cooling equations. For metal cooling in gas $T>10^4$\,K, we use tabulated cooling rates from {\small CLOUDY} \citep[][]{Ferland+98}. Below this temperature, cooling is dominated by fine structure lines, using cooling rates from \citet[][]{RosenBregman95}. We compute the radiation pressure exerted by photons, including contributions from both direct and reprocessed (infrared) radiation. % Dense gas clumps (e.g. molecular clouds) are self-shielded against ionising radiation, which is handled by the RT module. 

%%%%%%%%%%%%%%%%%%%%%%%%%%%%%%%%%%%%
\subsection{Post-processing with \rascas}\label{sec:rascas}
The \rascas code \citep[][]{Michel-Dansac+20} is a parallelised, post-processing, Monte Carlo code for propagating radiation through a 3D mesh. The code is aimed towards the complex scattering effects of resonant lines, such as \Lya, but can also handle other lines, as well as stellar continuum. The amount of photon packets emitted is user-defined, and their distribution between cells/stars is (in this work) weighted based on the source's luminosity. To perform mock observations, \rascas applies a peel-off algorithm \citep[see e.g.][]{Dijkstra17} to calculate the probability of a photon to escape along the chosen mock sightline at each scatter event.

% \citep[see e.g.][]{Yusef-Zadeh+84, ZhengMiralda-Escude2002, Dijkstra17} to calculate the probability of a photon to escape along the chosen mock sightline at each scatter event.

% \rascasalt works in three steps: construct the domain mesh used for propagation, calculate the initial photons packets, and then propagate those photons. Photon packets are propagated along a direction until it encounters a scattering or absorption event, or until it reaches the specified volume boundary. For a scattering event, the photons are be Doppler shifted and scattered to a new direction. 

In this work, we trace the propagation of LyC, \Lya, and \Ha. For the emission lines, we calculate the recombination and collisional emission from the gas using emissivity equations, as well as the continuum from the stars. To this end, we make use of the modules part of the base \rascas code that calculate the initial \Ha\ and \Lya\ photons in the gas from RT variables directly from a \ramses output. \Lya\ is a resonant-line and scatters with deuterium and neutral hydrogen, using line-parameters from the NIST\footnote{ \url{https://physics.nist.gov/PhysRefData/ASD/lines_form.html} } database. Dust extinction is treated by the SMC law\footnote{We briefly compared with using the LMC extinction law, but saw no significant change in our results.} \citep[taken from][]{Laursen+09} by assuming a dust-to-metal ratio which scales with metallicity in each gas cell, with a cut-off if the temperature goes above $10^4$\,K. The albedo and anisotropy coefficient of the dust were taken from Table 6 in \citet[][]{LiDraine2001} for each waveband respectively.

When cells are de-refined, the resulting parent cell is attributed a temperature and ionization fractions that are averages of the merged children cells. This can create unusual combinations of temperature and ionisation states that may produce artificially enormous \Lya{} or \Ha{} emissivities. In order to resolve this we filter the collisional emission in cells with a local cooling time below six timesteps. This choice of threshold ensures that most numerical artefacts are filtered away, but also inadvertently removes cells which are not artificially boosted. Nevertheless, we find that the impact on the total luminosity is relatively minor, often less than a factor of two (depending on the specific time output).

% Finally, when cells are de-refined the ionisation parameters are summed in such a way that the collisional component can be over-estimated; in order to resolve this we

Our implementation of LyC propagation is as follows. LyC is emitted by stars and its emission divided in three frequency bins, corresponding to the ionisation energy thresholds of {\rm H{\small I}}, {\rm He{\small I}}, {\rm He{\small II}} (same as for the RT module, Section~\ref{sec:RT}). LyC is attenuated by dust and H \& He, but can also be reflected by dust according to the albedo \citep[0.32; taken from][]{LiDraine2001}. The probability of an LyC photon to interact with H \& He is determined from their cross-section in the three different frequency bins. The number-weighted mean cross-section for each ion species is directly output by \ramsesrt and varies only slightly over time. 
%%% covering the wavelength range $0-912$\,Å.

%%% The effects of not resolving the cooling time of collisional Lya/Ha gas. "Need" to set a limit on cooling timescale to remove unresolved regions which are very bright in the collisional component.

% For \Lya, we traced radiation over the wavelength range 1200-1240\,Å, and for \Ha\ 6540-6600\,Å, but the range presented in figures varies according to what the focus is.

% In order to calculate the escape fraction later on, we also performed the exact same simulation but turning off dust interactions and setting the cross-sections to zero.

%%%%%%%%%%%%%%%%%%%%%%%%%%%%%%%%%%%%%%%%%%%%%%%%%%%%%%%%%%%%%%%%%%%%%%%%%%%%%%%%%%%%%%%%%%%%
%%%%%%%%%%%%%%%%%%%%%%%%%%%%%%%%%%%%%%%%%%%%%%%%%%%%%%
\subsection{Simulation suite -- Haro 11}
The simulations presented in this paper are based on the Haro 11 simulations first presented in E25. For context, we show an image of Haro 11 in the left panel of Figure~\ref{fig:Haro11_cont}. In this image, the most recognisable features of the galaxy are the three stellar knots (A, B, C) and the dusty gas arm, nicknamed the "ear" (outline in the right image). In E25 we determined that, in our simulations, knot B/C corresponds to the central bulge of the more/less massive progenitor galaxy, respectively. The more massive galaxy is gas rich, and its disc arm gives rise to the dusty ear feature and to knot A.

% We ran \rascas for three different wavebands and convolved the spectra with the same HST filters as in the observations. 

In E25 we focused on reproducing features of the Haro 11 galaxy, such as: the \emph{single} tidal tail, three stellar knots, and general morphology and kinematics. In this work we build on E25, focusing on the escape of radiation from the inner parts of our simulated Haro 11. We employ the same initial conditions, but with on-the-fly radiative transfer and at a spatial resolution four times higher (i.e. the minimum cell width is $\sim 9$\,pc, compared to $\sim 36\,$pc). % The simulations were improved upon by eye, to mach various features of Haro 11.

%%%%%%%%%%%%%%%%%%
% Next, we briefly outline the method used to derive the initial conditions for the simulations presented in this work, but a more detailed explanation and observational comparison can be found in E25.
The simulation setup is of a merger between two similar disc galaxies, in terms of mass and size. One of the galaxies is slightly more massive, more gas rich, and has a prograde rotation relative the interaction point. The other galaxy has a retrograde rotation, which inhibits the formation of a (second) tidal tail. The final setup was determined by iteratively tweaking the initial conditions to best match a variety of properties; morphology, kinematics, and stellar population features.

The initial masses of the two galaxies were calculated from observations of the stellar and gas (ionised, neutral, and molecular) mass. The progenitors are disc galaxies following a \citet{NavarroFrenkWhite96} profile for the dark matter halo ($M_{\rm DM}\sim 10^{11}\Msol$), a \citet{Hernquist1990} profile for the stellar bulges ($M_{\rm b}\sim10^{9}\Msol$), and an exponential distribution for the gas and stellar discs ($M_{\rm d}\sim10^{9}\Msol$, for both gas and stars). Each component contains $10^6$ particles for the DM and stars, except the stellar bulges which have $10^5$ particles. The full simulation volume is 300$^3$\,kpc$^3$. The orbital parameters are given in E25.

%%%%%%%%%%%%%%%%%%%%%%%%%%%%
% The progenitors are disc galaxies following a \citet{NavarroFrenkWhite96} profile for the dark matter halo ($M_{\rm DM}=[2\times10^{11},\, 1.67\times10^{11}]\Msol$), a \citet{Hernquist1990} profile for the stellar bulges ($M_{\rm b}=[4.00\times10^{9},\, 2.28\times10^{9}]\Msol$), and an exponential distribution for the gas and stellar discs ($M_{\rm d,g}=[5.5\times10^{9},\, 1.2\times10^{9}]\Msol$ and ($M_{\rm d,*}=[4.57\times10^{9},\, 2.57\times10^{9}]\Msol$)).
%%%%%%%%%%%%%%%%%%%%%%%%%%%%%%%

%%%%%%%%%%%%%%%%%%%%%%%%%%%%%%%%%%%%%%
%%%%%%%%%%%%%%%%%%%%%%%%%%%%%%%%%%%%
\subsection{Constructing mock observations}\label{sec:mock_method}
Here we outline the method in which we analyse our simulations from multiple outputs and then explain how \rascas is applied to produce our mock observations. Our simulations output data every 5\,Myr during the merger, starting $\sim50$\,Myr before the galaxies first interact until present-day, when the merger most closely resembles Haro 11 (set to be at 0\,Myr in this work). To better capture this final stage, outputs are produced every 1\,Myr. We use the position of the bulge of the more massive galaxy as the central point to follow throughout the merger. SFRs are calculated by summing the stellar mass that formed within 5\,Myr bins and then dividing by this bin size.

The \Lya\ and \Ha\ profiles are both combinations of a stellar continuum, collisional and recombination emission (see Section~\ref{sec:rascas}). The profiles and values for \Lya\ presented in this work were derived after fitting and subtracting the continuum. The purpose of performing a fit (when the continuum is already known) is to capture any absorption feature. Including the spectral absorption is commonly done in observations, since it is connected to the underlying physics of \Lya\ propagation. Finally, angle-average escape fractions are calculated by counting the escaping photons from the simulation volume, while mock observation (directional) escape fractions are defined from the intrinsic/escaping luminosity as
% , which often shows clear absorption feature on the blue side
\begin{align}
    f_{\rm esc} = \frac{L_{\rm escaping}}{L_{\rm intrinsic}}.
\end{align}
This differs from observations, which often have to use proxies to calculate the intrinsic emission.

For \rascas mock observations of the galaxies over time, we defined 100 different angles distributed evenly around the galaxies, centred on the more massive galaxy. We measured the spectra along each direction within an aperture of 150\,kpc, using a spectral resolution of 0.1\,Å. For imaging of the galaxy at present-day, we used spatial resolutions of 50\,pc. Data cubes used a spatial resolution of 100\,pc and 0.1\,Å for the spectra. All of the \rascas post-processing was done with $10^6$ photon packets, with the exception of \Lya\ which used $10^5$ photons due to the extensive computational time required to track resonant scattering. Imaging of \Lya{} used $10^6$ photons.
%This allows us to get an understanding of the variation due to the angle of observation.

For direct comparison with observations of Haro 11, we define the position and range of the three knots (A, B, C) as seen on the right image in Figure~\ref{fig:Haro11_cont}, circled in red. The diameter of the aperture is 1\,kpc for knot B \& C, while knot A has an irregular region with a slightly larger size, to account for its more extended structure. All mock observations visualised in this paper are from the same angle of observation. This sightline yields the best match observed morphology and kinematics (see E25 for details).
% The region for Knot A is more extended, which was done to better encompass all the stars scattered around the knot. 

The mock observation image of Haro 11, on the right in Figure~\ref{fig:Haro11_cont}, was constructed using \rascas{}. Tracing the continuum, we produced three data cubes that covered the same three waveband ranges as the HST composite image on the left image. The mock spectra were then convolved with the same HST filters \citep[220W, 435W, 814W;][]{Adamo+10} and integrated.

%%%%%%%%%%%%%%%%%%%%%%%%%%%%%%
%We calculate the mock escape fraction as is commonly done in the literature \citep[e.g.][]{Ostlin+21}
%\begin{align}\label{eq:Lya_intr}
%    f_{\rm esc}^{\rm Ly\alpha} = \frac{L_{\rm Ly\alpha}}{L_{\rm Ly\alpha, int}},\ \ L_{\rm Ly\alpha, int} = 8.4\times L_{\rm H\alpha,\,no\ dust}
%\end{align}
%where $L_{\rm Ly\alpha, int}$ is the intrinsic \Lya\ luminosity, and $L_{\rm H\alpha,\,no\ dust}$ is the dust-corrected \Ha\ luminosity.

\begin{figure*}
    \includegraphics[width=0.95\textwidth]{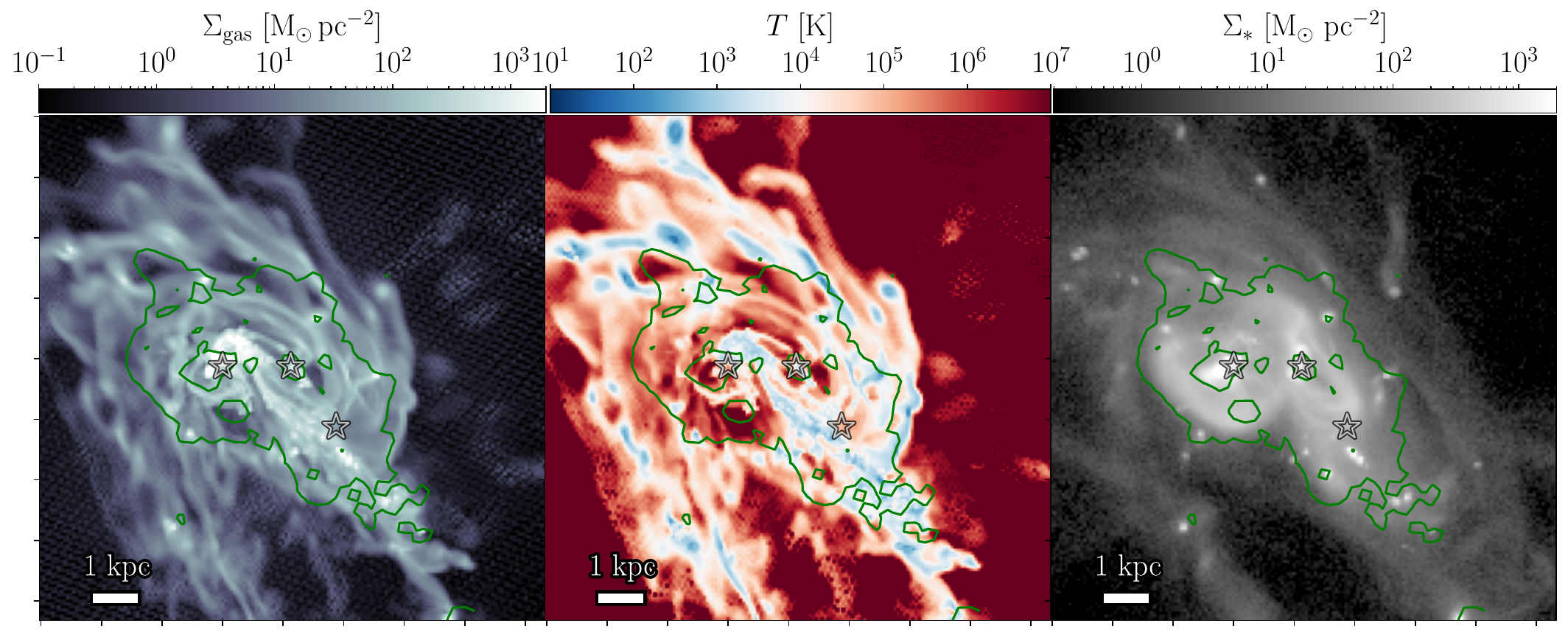}
    \caption{ From left to right: The projected gas surface density, gas mass-weighted mean temperature, and the stellar surface density of the galaxy at present-day. The green contour corresponds to the (attenuated) stellar continuum in Figure~\ref{fig:Haro11_cont}, and the star-symbols are the three knots. }
    \label{fig:dens_temp_star}
\end{figure*}

\begin{figure*}
	\includegraphics[width=0.95\textwidth]{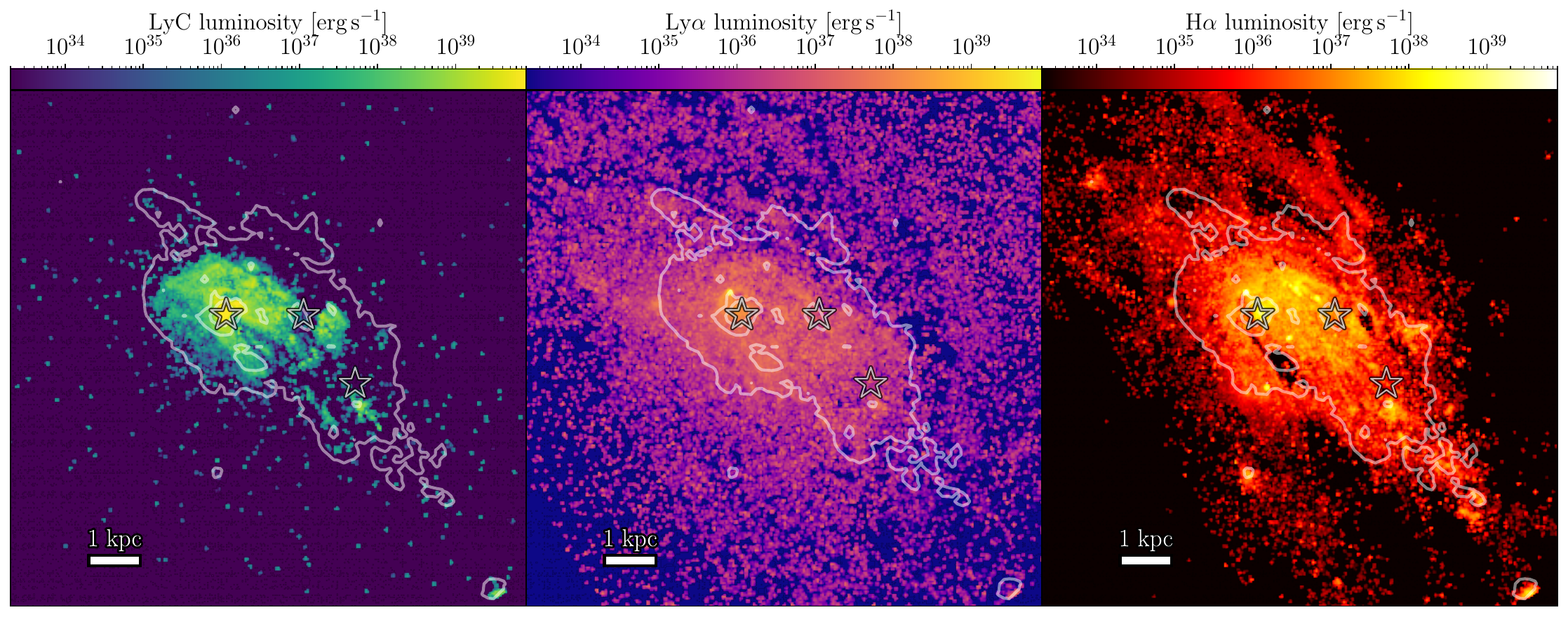}
    \caption{ The escaping luminosities of LyC, \Lya, and \Ha, respectively. Annotated as a white contour is the (attenuated) stellar continuum and the position of the knots are represented by star-symbols (see Figure~\ref{fig:Haro11_cont}).  } %% These are images that include the continuum
    \label{fig:triple_mock_image}
\end{figure*}

\begin{figure}
	\includegraphics[width=0.48\textwidth]{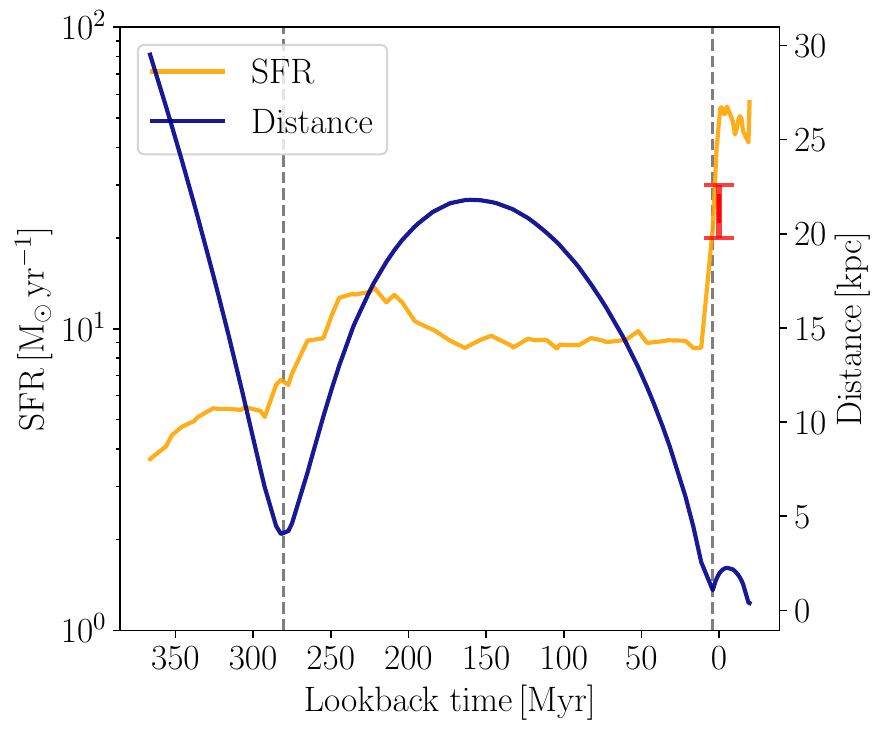}
    \caption{ The distance between the two galaxies and the total SFR as a function of time. The time for present-day Haro 11 is defined to be at 0 Myr. The two pericentres are marked as dashed lines. The approximate range of the SFR from observational data \citep[$\sim20-30\Msolyr$;][]{Hayes+07, Adamo+10, Madden+13, MacHattie+14, Gao+22} is marked with a red errorbar at present-day. }
    \label{fig:sfr_distance_time}
\end{figure}

%%%%%%%%%%%%%%%%%%%%%%%%%%%%%%%%%%%%%%%%%%%%%%%%%%
\section{Results}\label{sec:results}
% In E25, we identified that
\subsection{Visualising the gas, stellar, and emission morphology }\label{sec:mock_obs}
First, we visualise the physical condition of the ISM within our simulated galaxy at present-day. The projected gas surface density, gas temperature, and stellar surface density, are shown in Figure~\ref{fig:dens_temp_star}. The gas morphology exhibits a dense clump between knots B and C, which is likely due to compression from the recent, close interaction. A cold, dense extension to the right of knot B extends upwards and above it, which we identify as the 'ear' observed in Haro 11. This feature origins from the disc around knot B, which also provides the gas around knot A. The stellar density is at its highest around knot B \& C, which are the progenitor galaxies' stellar bulges. Knot A appears more dispersed, i.e. less dense, but still contains clusters of bright stars.

% The stellar and gaseous structures of the simulated galaxies are presented in detail in E25. In this paper, we will visualise the galaxy through mock observations.

Next, a mock observation of the escaping emission of LyC, \Lya, and \Ha\ from our simulated galaxy is presented in Figure~\ref{fig:triple_mock_image}. A white contour of the stellar continuum from Figure~\ref{fig:Haro11_cont} is plotted on top to help orient in the image. The LyC emission is brightest around knot C, with some radiation also escaping close to the other two knots. Same is true for the \Lya\ map, although there is also a more diffuse component surrounding both knot B \& C. Knot A contains two distinct sources of \Lya, but also a diffuse component extending from knot B. The \Ha\ emission follows a similar distribution, but is more extended and brighter around knot B. Notably, the bright clumps in \Ha{} highlight H{\small II} regions, which are forming from ionised gas around newborn stars, while the extended gas is emitted by the diffuse ionised gas phase \citep[see][]{Ejdetjarn+24}. This is only the radiation escaping the galaxy, and we will compare this to the intrinsic emission in context of the spatially resolved escape fraction in Section~\ref{sec:spatially_resolved_fesc}.

%%%%%%%%%%%%%%%%%%%%%%%%%%%%%%%%%%%%%%%%%%%%%%%%%%
\subsection{SFR, luminosity, and escape fraction during the merger}\label{sec:escape_fraction}
Our merger simulation of the Haro 11 galaxy undergoes two close interactions before reaching the state considered to be the present-day Haro 11 (see E25). In Figure~\ref{fig:sfr_distance_time}, we show the SFR and the distance between the two progenitor galaxies over time. The pericentre of the two interactions, marked by vertical dashed lines in grey, induce starburst episodes; the first period is delayed by a few tens of Myr, while the latter is more immediate due to the closer interaction leading to a more direct collision. The range of observational data is shown as a red marker. At the present-day, 0\,Myr, the galaxies have just passed each other ($3-5$\,Myr after pericentre) and the SFR has risen to $\sim 40-50\Msolyr$. 

The intrinsic luminosity, escaping luminosity, and escape fraction of LyC, \Lya, and \Ha{} are presented in Figure~\ref{fig:luminosities}. The grey dashed vertical lines indicate the pericentres. The general appearance of the intrinsic luminosity is very similar between the three emissions, as they closely follow the SFR history. On the other hand, their escaping luminosities have similar values but evolve differently throughout the merger. Notably, the emergent LyC increases by an order of magnitude at the pericentres, while \Lya{} and \Ha{} show a shallow increase. The escape fractions further highlight these differences: LyC exhibits sharp increases at both pericentres and at the SFR peaks, whereas \Lya{} and \Ha{} have lower escape fractions during these peaks, explaining their weaker escaping emission. Additionally, as the galaxies move closer, the \Lya{} and \Ha{} escape fractions decrease slightly and remain lower than their initial values. We discuss reasons for this behaviour, as well as the impact of sightline variation, in Section~\ref{sec:fesc_angle_var}. Near the present-day, the LyC and \Lya{} escape fractions fall within the observed range of values for Haro 11, shown as shaded regions. 

%  From the escape fractions, we note that LyC exhibits a sharp increase at both of the pericentres and at the SFR peaks. Meanwhile, the \Lya{} and \Ha{} escape fractions remain largely unchanged at the SFR peaks, explaining the lower escaping emission of these lines. 

% First, consider the \emph{total} luminosity and escape fraction of LyC, \Lya, and \Ha, which are the actual values for the escaping radiation; derived directly from \rascas by counting the photons escaping the simulation volume.

% The total escaping luminosity and escape fraction of LyC, \Lya, and \Ha, over the duration of the merger is shown in Figure~\ref{fig:lum_fesc_angle_average}. This data is derived directly from the \rascas runs by calculating the luminosity and fraction of the photons that escape the simulation volume, which is different from a mock observations with a set viewing direction and aperture.

\begin{figure}
	\includegraphics[width=0.45\textwidth]{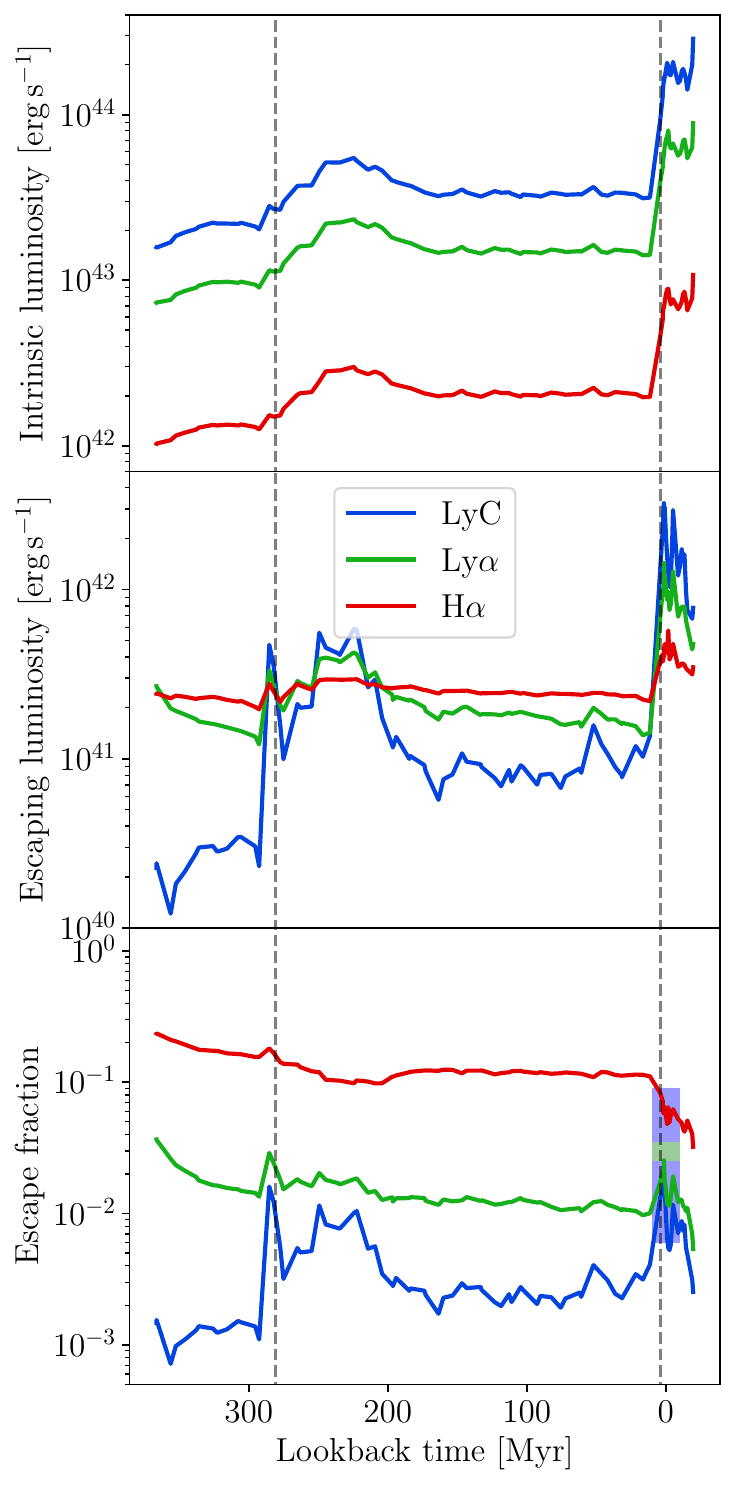}
    \caption{ The intrinsic luminosity (top), escaping luminosity (middle), and the angle-average escape fraction (bottom). The pericentres of the merger are shown as vertical, dashed lines. Present-day is at 0\,Myr. The shaded regions in the bottom plot correspond to the observed escape fractions of LyC \citep[blue; ][]{Bergvall+06, Grimes+07, Leitet+11, Komarova+24} and \Lya{} \citep[green;][]{Hayes+07}. }
    \label{fig:luminosities}
    % The escape fraction is calculated from the total number of photon packets escaping the volume over the number of packets emitted. 
\end{figure}

\begin{figure}
    \includegraphics[width=0.45\textwidth]{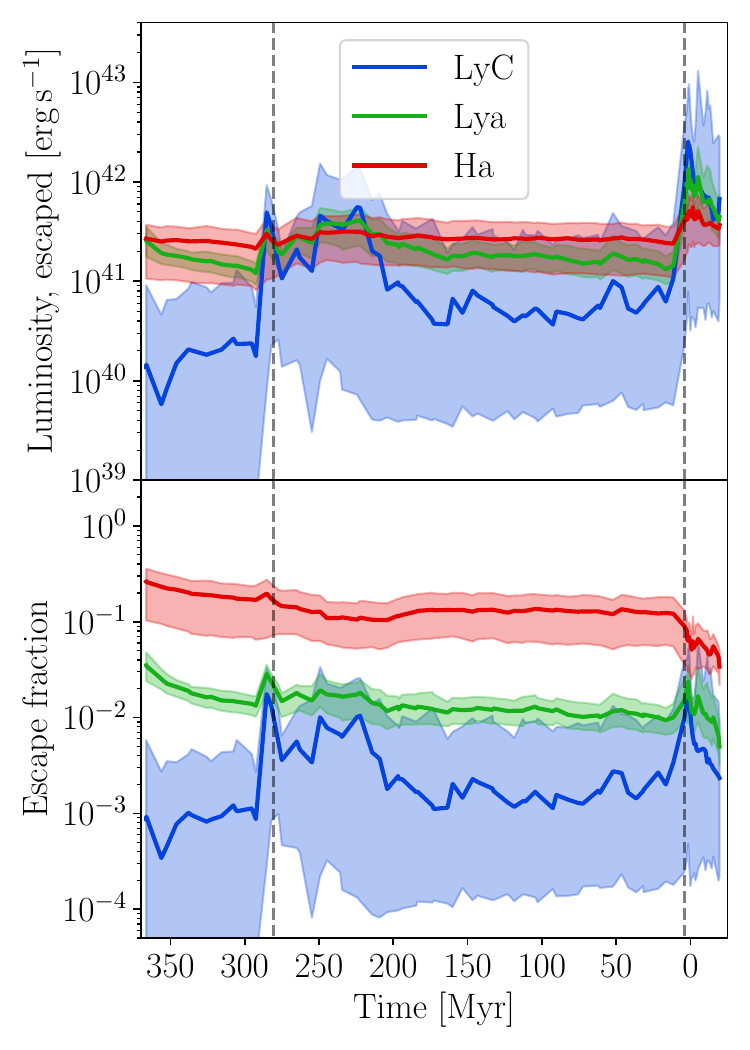}
    \caption{ The attenuated luminosities and escape fractions from mock observations of LyC, \Lya{}, and \Ha{}. The lines show the medians, and the shaded regions correspond to the 95 percentile of the data and represents the variation in different sightlines. The pericentres of the merger are shown as vertical, dashed lines.  }
    \label{fig:lum_fesc_mock}
\end{figure}

\begin{figure*}
    \centering\includegraphics[width=0.85\textwidth]{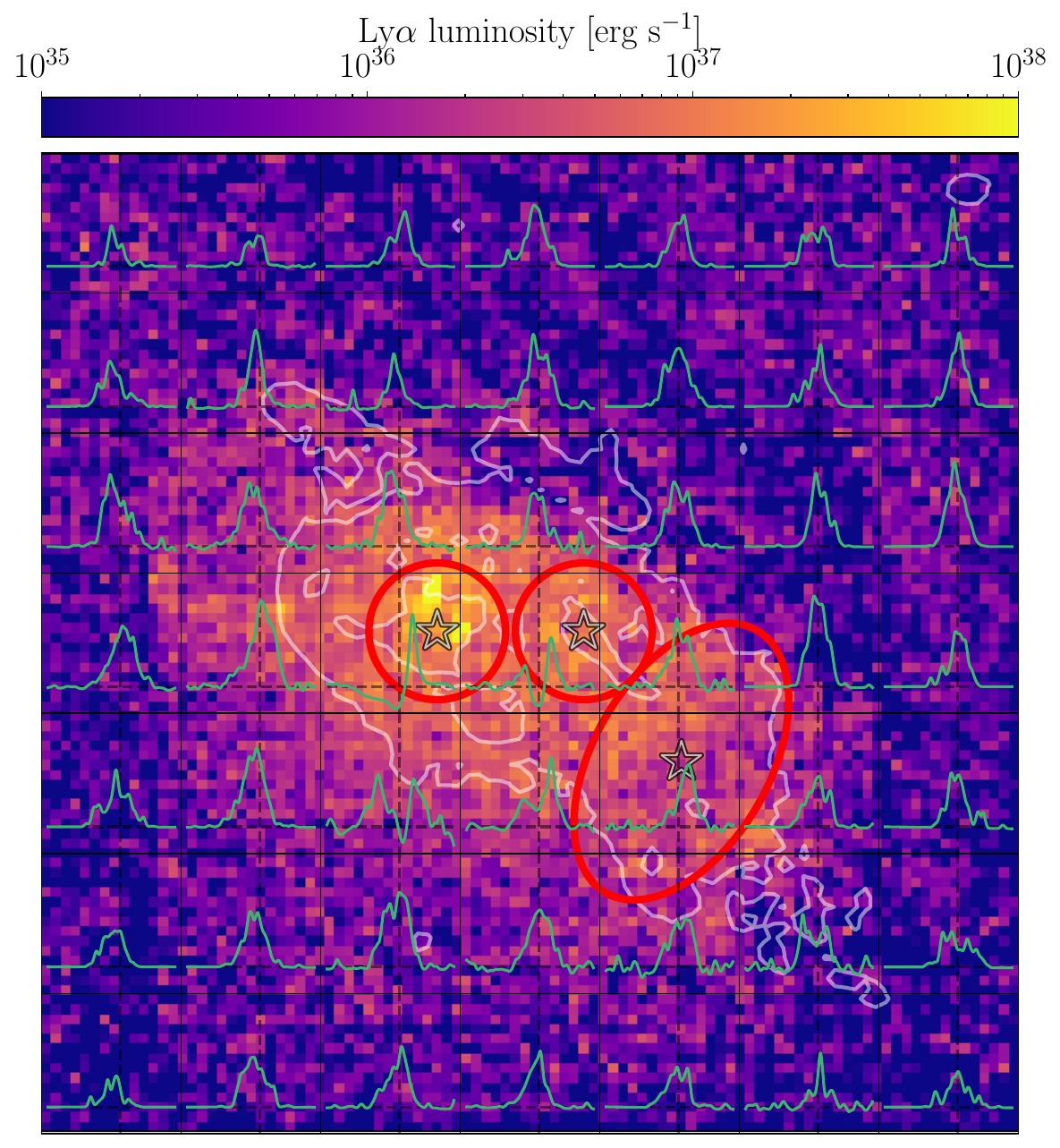}
    \centering\includegraphics[width=1.0\textwidth]{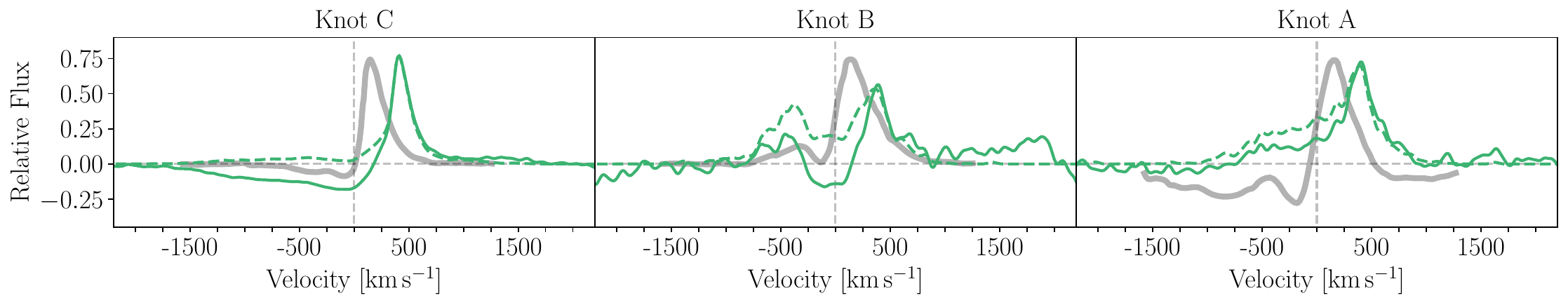}
    \caption{ \emph{Top:} A projection map of the attenuated \Lya{} luminosity with a grid of the \Lya{} line profiles annotated. The map was produced from a spectral cube by fitting the \Lya\ flux in each pixel and integrating over it. The white contour represents the (attenuated) stellar continuum from Figure~\ref{fig:Haro11_cont}. The profiles are normalised to the local peak value and shown for the velocity range [$-2\,000$\kmsec$, +2\,000$\kmsec]. Each grid cell has a grey, dotted vertical/horizontal line that highlights the zero point for the velocity/luminosity. \emph{Bottom:} The \Lya{} lines in the three knots, which are roughly defined by the red circular contours. The green solid line is the escaping emission, while the dashed line is the escaping emission without the contribution from the continuum. The grey profile is the observed \Lya\ profile from \citet[][]{Ostlin+21}, normalised for comparison. }
    \label{fig:lyagrid}
\end{figure*}

%%%%%%%%%%%%%%%%%%%%%%%%%%%%%%%%%%%%%%%%%%%%%%%%%%
\section{Mock observations}\label{sec:discussion}

%%%%%%%%%%%%%%%%%%%%%%%%%%%%%%%%%%%%%%%%%%%%%%%%%%
\subsection{Angular variation of luminosity and escape fraction}\label{sec:fesc_angle_var}
The luminosities and escape fractions from the mock observations of the galaxy during the merger are presented in the upper panel of Figure~\ref{fig:lum_fesc_mock}. The mocks were constructed by tracking emission escaping the galaxy from 100 directions, distributed uniformly on a sphere centred on the galaxy, and at each timestep (see Section~\ref{sec:mock_method} for details). The lines show the median and the shaded regions are the 95 percentile, represents variation due to sightlines. The pericentres of the merger are shown as grey, dashed lines. As expected, the median luminosities follow the overall shape of the angle-averaged luminosities in Figure~\ref{fig:luminosities}. 

% resonance scattering in combination with dust extinction 

% However, the median \Lya{}luminosities and escape fractions are slightly higher at certain points (e.g. around the peak around 250 Myr and the second pericentre) which might indicate that 100 beam directions do not fully capture the angle-average or that we are underestimating the continuum subtraction. This is further discussed in Section~\ref{sec:caveats}.

In the lower panel of Figure~\ref{fig:lum_fesc_mock}, we show the escape fraction of LyC, \Lya, and \Ha{} as a function of time, which also is in agreement with the angle-average in Figure~\ref{fig:luminosities}. These escape fractions are slightly lower than expected from observations. In particular, an $f_{\rm esc}^{\rm H\alpha}\sim10\,\%$ indicates that there is a lot of dust absorption, which we discuss in Section~\ref{sec:caveats}. Furthermore, the LyC escape fraction can be seen to increase as the galaxies are close, and reach a peak at the pericentres. A broader peak appears directly after the first pericentre, which is in synch with the SFR burst at this time (see Figure~\ref{fig:sfr_distance_time}).

The increase at the first pericentre is very sharp and not associated with any strong peak in SFR, which makes feedback unlikely to be the main contributor. Instead, this likely occurs due to dynamical effects, such as tidal or ram pressure stripping of neutral gas. The latter of these relies on the collisionless nature of the stars: when the galaxies interact the gas is affected by ram pressure and stars can be displaced into low-density regions. This idea is visually in agreement with Figure~\ref{fig:dens_temp_star} at the second pericentre, as the most prominent gas and stars clumps (around knot C) appear to be spatially offset. Similar displacement has been seen in recent ALMA between dust and star-forming regions \citep[e.g.][]{Inami+22}. In the same vein, \citet[][]{KostyukCiardi2024} partially attributed LyC leakage to the displacement of neutral gas. Furthermore, compressive tidal forces could be producing a more clumpy ISM \citep[][]{Emsellem+08, Renaud+21a}, which results in regions of low-density gas outside these clumps \citep[reminiscent of the "picket-fence" model, see e.g.][]{Rivera-Thorsen+17b}.

However, stellar feedback could still explain the increase in LyC escape fraction at the pericentres (by forming escape channels in the dense gas), but as the peak is very sharp and the starburst period has only barely begun, it would require intense feedback from young ($<5$\,Myr) stars (discussed further in Section~\ref{sec:caveats}). In contrast, the higher escape fraction following the first pericentre ($t=200$–$250$ Myr) coincides with a prolonged SFR burst and can therefore be more readily explained by standard feedback mechanisms.
%Although, the higher escape fraction after the first pericentre, $t=200-250$\,Myr, coincides with an SFR burst and can thus be more readily explained by various feedback mechanisms. 

% favourable geometrical setup as the luminous progenitor galaxies' bulges (knot B \& C) move closer and sightlines overlap. The increase could also be due to
% (overlapping their sightlines), since the bulges are the densest star forming regions, feedback effects are more frequent

% that has more sightlines with low gas density, which favour the escape of ionising radiation. These sightlines could occur from an overlap in escape channels 

Similarly, the \Lya\ follows the same overall shape as the LyC but with much smaller amplitudes and variation. The combination of LyC escape increasing and \Lya\ escape not varying much is interesting, as ionising radiation is an integral part in the production of \Lya, and it has been suggested that the presence of one could be used to infer the other \citep[e.g.][]{Verhamme+15, Izotov+21, LeReste+25c}. Conversely, the escape fraction of \Ha\ shows only a shallow increase at the first pericentre and a decline during the second one. The overall low \Ha{} escape fraction is due to dust absorption, and could reflect that young, luminous, stars are born embedded in dusty regions from which their intrinsically bright radiation is unable to escape. The decrease in \Ha\ escape fraction during the SFR peaks implies enhanced absorption, which could be either due to higher gas densities and a clumpier ISM during the merger (i.e. stars more heavily embedded) or an increased metallicity enrichment from supernovae (causing higher dust production).

Furthermore, the same process of stellar displacement could be responsible here as well; LyC is emitted from stars, while \Lya{} and \Ha{} are emitted from the gas. The gas environment during these close passages exhibits an increased gas density and a decrease of ionisation within the clouds, since the stars are displaced and no longer embedded in the gas. This results in dense, dusty clouds that have a higher absorption rate for nebular emission. As stars and gas again coalesce, they follow a similar behaviour and all emission types are similarly hampered after the second pericentre passage. Additionally, \Lya{} remains strongly scattered at low \HI{} column densities that are transparent to LyC, owing to its resonant interaction with neutral hydrogen, meaning that a LyC escape channel does not necessarily permit \Lya{} escape.
% Additionally, \Lya{} remains strongly scattered at low \HI{} column densities that are transparent to LyC (due to its much higher cross section), meaning that a LyC escape channel does not necessarily permit \Lya{} escape.

The variation in escape fraction (and luminosity) with observation angle in \Lya\ and \Ha\ is relatively minor. Moreover, the distribution of \Lya{} luminosity is expected to be broader due to dust absorption, which causes strong variation between lines of sight \citep[as demonstrated in disc galaxy simulations;][]{Smith+22, Blaizot+23}. However, collisional emission has a significant contribution to the total escaping signal in our simulation ($\sim20-40\,\%$, depending on snapshot) and is prominent in the outskirts of the galaxy, from where it escapes without much extinction. This limits the variation in emergent \Lya{} emission.

Meanwhile, LyC varies by about two orders of magnitude between observations along different sightlines. This is due to its high photoionisation cross-section with dust and gas, which means LyC requires paths clear of both dust and gas in order to escape. At the first pericentre, even the lines of sight that normally exhibit low LyC escape (lower shaded region) show a pronounced increase. This indicates that the enhancement is not driven by a localised event, such as a supernova opening a narrow channel, but instead reflects global restructuring of the gas that reduces obscuration across many directions at once. The mechanisms that cause this are likely the same that drive the increase in the median LyC escape fraction, as discussed previously in this section.

% In particular, at the first pericentre even the line-of-sights with low LyC escape fraction show a drastic increase. This further confirms that there is some underlying physical processes that clears gas in multiple directions at once, not just a single supernova clearing gas along a few line-of-sights. This can likely be attributed to the same effects that increase the median LyC escape fraction (discussed above), such as a brief displacement between stars and gas during the close encounter.

% We hypothesise that an increase in the gas density during the merger, leads to more \Lya\ photons being absorbed, while the LyC absorption is already saturated along these sightlines.

%%%%%%%%%%%%%%%%%%%%%%%%%%%%%%%%%%%%%%%%%%%%%%%%%%
\subsection{The \Lya\ spectra throughout the galaxy}\label{sec:lya_spectra}
The \Lya\ line profile can take on many different shapes depending on the physical conditions of the gas and dust it interacts with. Observations of Haro 11 by \citet[][]{Ostlin+21} explored the \Lya\ line for different regions of the galaxy. We qualitatively compare the \Lya\ lines in our simulations with these observations by dividing the galaxy into a grid of $1$\,kpc-wide squares (roughly matching the aperture of the \Lya\ observations in \citealt{Ostlin+21}) and sum up the total spectra within each square. This grid is plotted over a projection map of the \Lya\ emission in Figure~\ref{fig:lyagrid}. In order to take into account the \Lya\ absorption feature, we make a fit to the continuum, remove it from the total spectra, and then integrate over the resulting \Lya\ line. This is done for each pixel. The same fitting method is used for the over-plotted line profiles.
% Compared to the \Lya\ map in Figure~\ref{fig:triple_mock_image}, this map does not include the continuum. 

% However, extragalactic observations identify stars in clusters and the stellar and age distributions are thus different from the individual star particles in our numerical model. Further analysis to identify clustering in our simulations is outside the scope of this work.
% A quantitative comparison with observations is useful evaluate the physical conditions of the observed galaxy, if the observed properties match. 

From this figure we first note that the \Lya\ line varies within the galaxy, and can be quite different between neighbouring cells, meaning that a slight shift of $\lesssim0.5$\,kpc can result in an absorption/emission feature being added or removed. This is due to sharp changes in the gas conditions between some grid cells. Most of the profiles outside of the galaxy are at their rest-frame wavelength, as there is little neutral gas and no significant out/inflows this far out. A few strong double peaks are apparent, which can be interpreted as low neutral H content for \Lya\ to scatter on, and with little dust extinction. The escaping \Lya{} emission in the outskirts of the galaxy is dominated by collisional excitation, while the emission around the knots is mainly from recombination; however, they contribute roughly equally to the total escaping luminosity at present-day.

% The collisional component is more homogeneously spread out and, in particular, make up most of the emission outside of the galaxy, where dust extinction is much less present. % {This pattern has been reported in simulations of merging or high-z star-forming galaxies (e.g. Verhamme et al. 2006; Smith et al. 2019; Behrens et al. 2018), where resonant trapping plus dust extinguish the recombination core emission but shock- or wind-induced collisional emission leaks out more efficiently.}

Furthermore, there is a strong emission peak around knot C, which we have previously shown is the dominant source of escaping LyC and \Lya\ emission, in agreement with observations \citep[][]{Ostlin+21}; \citet[][]{Komarova+24} also show knot C is a strong emitter, but find knot B to be dominant. The spectrum here shows a strong, redshifted emission line, and some absorption on the bluer side. On the cell to the right (between knot B \& C) there is noticeable absorption on the blue side of the spectra, which roughly overlaps with a region of high gas density (see left plot of Figure~\ref{fig:dens_temp_star}, just to the right of knot C).

As the three stellar knots are of particular interest, we plot the spectra around these separately, underneath the map in Figure~\ref{fig:lyagrid}. The dashed green lines show the escaping spectra without the absorption feature (i.e. continuum excluded) and the grey lines show real observational data from \citet[][]{Ostlin+21}. The knots positions are roughly defined by the red circles, which do not fully overlap with any single grid cell but are rather a combination of the surrounding cells. For example, the strong absorption in knot B corresponds to the absorption seen in the middle of the map. Just to the left of this cell, the strong peak and shallow absorption features of the \Lya\ profile in knot C are visible.

Compared to observations (grey lines), the mock spectra show qualitative similarities with various features. The redshifted \Lya\ line in knot C indicates that the emission has undergone several scatterings with neutral gas and/or gas inflow. The lack of a blue peak and the absorption feature implies that there is dust present. Knot B shows two clear peaks, although with a notable dust absorption as well. Knot A also has two peaks, but the observed spectrum indicates there should be more dust than our mocks show. Additionally, all of these spectra are slightly more red-shifted than the observations. This might be due to the systematic velocity of the galaxy being difficult to define, since the line-of-sight velocity varies widely for the knots (progenitor galaxies); $v\approx(+60, -60)$\kmsec for knot B and C, respectively. It could also be due to our simulation exhibiting either faster gas kinematics or higher \HI{} content, as the latter would force the \Lya{} line to undergo more scatter and shift in phase space.
%%%  which might be a difference in removing the systematic velocity

\begin{figure*}
	\includegraphics[width=0.95\textwidth]{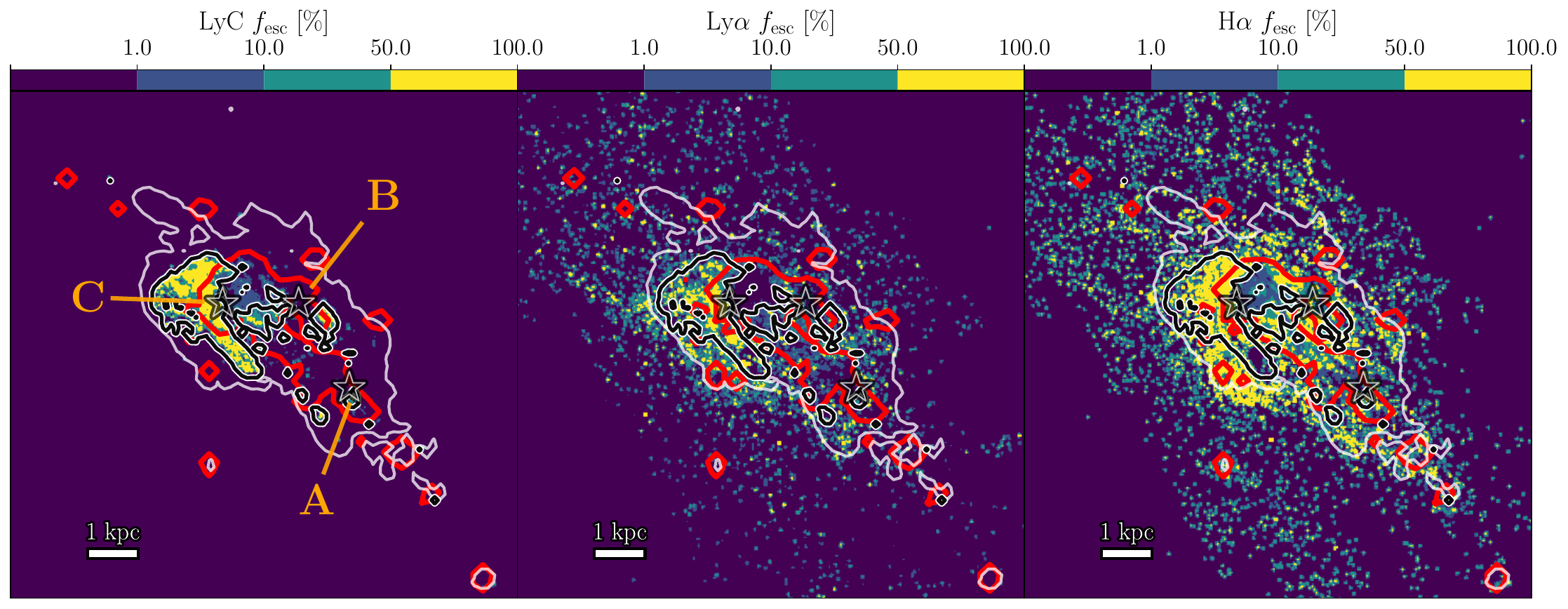}
    \caption{ Resolved maps of the escape fraction in LyC, Ly$\alpha$, and H$\alpha$. The white contour is the outer contour from the stellar continuum map in Figure~\ref{fig:Haro11_cont}. The black contour with a white outline corresponds to the LyC escape fraction (leftmost map). The star-symbols are the approximate positions of the knots. The red contours represent the intrinsic emissivity from the respective sources. Pixels with no intrinsic/emergent luminosities are set to a lower limit based on the minimum values of their respective maps. } %  The intrinsic Ly$\alpha$ is calculated from equation~\ref{eq:Lya_intr}.
    \label{fig:fesc_map}
\end{figure*}

%%%%%%%%%%%%%%%%%%%%%%%%%%%%%%%%%%%%%%%%%%%%%%%%%%%%%%%%%%%%%%%%%%%%%%%%%%%%%%%%%%%%%
\subsection{Spatially resolved escape fraction}\label{sec:spatially_resolved_fesc}
We next investigate where LyC, \Lya, and \Ha\ radiation is produced and where they escape by creating resolved maps of their escape fractions. Figure~\ref{fig:fesc_map} shows the escape fraction of each emission, with white contours of the stellar continuum, red contours corresponding to 95 percentile of the intrinsic emission, and black contours highlighting LyC escape. Thus, we use these maps to understand the general escape distribution, rather than absolute amount of radiation.
% The escape fractions can be above unity in specific pixels due to stochasticity of the Monte Carlo method and scattering, for which the latter is especially relevant for \Lya{}.

%  As LyC emission is exclusively from stars in our simulation, and given its high likelihood of destruction, we would not expect it to scatter far away from its site of production. 
From the leftmost panel we note that the LyC is primarily escaping around knot C, which matches fairly well with our map of the escaping LyC luminosity (Figure~\ref{fig:triple_mock_image}) and is slightly offset from the main body of its bright intrinsic radiation (red contour). LyC emission is exclusively from stars in our simulation and has a high likelihood of destruction due to large cross-section with gas and dust (although, with a $\sim30\,\%$ chance of dust scatter at each encounter; \citealt{LiDraine2001}). Thus, this radiation is directly escaping from sources just outside of knot C that are not obscured by dust and gas, with only a minor contribution from dust scattering.  

In the middle panel, the escaping \Lya\ roughly overlaps with the region of LyC escape, and the regions of bright intrinsic emission are also rather similar. However, the escaping \Lya\ is more extended and, thus, is scattering further away from its source. One interpretation of this difference is that stars are displaced relative to dense gas. Then, stars still ionize the surrounding lower-density medium and produce \Lya{} emission, but the regions of LyC emission and \Lya{} production would be spatially offset.

In the right plot, we show that the \Ha\ emission extends even further out. The escape fraction varies a lot through different parts of the galaxy, and most of the \Ha\ outside of the galaxy is escaping, which is generally found in observations. As \Ha\ is only absorbed by dust, the escape fraction map can be used as a proxy for the dust absorption. Applying this logic, we confirm there is a significant amount of dust in central parts of knot B \& C and in the 'ear' feature extending down to knot A and above knot B. Furthermore, the position with brightest intrinsic production coincides with a region of very low escape fractions (i.e. high dust content) to the right of knot C, which contains high-density gas (see Figure~\ref{fig:dens_temp_star}). This demonstrates that most of the intrinsic emission is embedded in dense, dusty clouds and gives an explanation for the low overall $f_{\rm esc}\sim 10\,\%$ for \Ha, seen in e.g. Figure~\ref{fig:lum_fesc_mock}.

% If there is a significant displacement between the stars and the gas, then scattering is not necessary; as the gas is disconnected from the ionising (LyC) source.

% The intrinsic \Lya\ is similarly distributed as the intrinsic LyC but the escaping radiation scatters further away from its source.
% Curiously, there is escaping \Lya\ emission just below knot A which is not present in the LyC

The luminosities and escape fractions from the three individual knots, as outlined in Figure~\ref{fig:Haro11_cont}, are presented in Table~\ref{tab:f_esc}. The LyC escape fractions are roughly the same in each knot, but the escaping luminosity is an order of magnitude higher for knot C. Thus, knot C is also producing about ten times as much ionising radiation as knot B or knot A. Compared to LyC observations of Haro 11 by \citet[][]{Komarova+24}, our knot B and C are 1-2 orders of magnitude more luminous. This can be explained by a difference in spectral range, as the authors were restrained to a limited window (903-912\,Å). By integrating our spectra over their range, we find a closer match. While we only make a qualitative comparison, an agreement with the observed $L_{\rm LyC}$ is encouraging for our model and the adopted sightline; as we have already shown, $L_{\rm LyC}$ varies by orders of magnitude between different sightlines.

Additionally, \citet[][]{Komarova+24} found that knot B is the dominant source of escaping and intrinsic LyC emission, but is obscured by dust and suffers more absorption than knot C. Thus, they report knot C has a higher escape fraction, despite having a lower observed LyC. In our simulations, knot C has a higher intrinsic \emph{and} escaping luminosity, but comparable escape fraction to knot B. Although, the escape fraction map shows that the region left of knot C is more prone to LyC leakage. The low escape fractions are connected to the gas density distribution, see Figure~\ref{fig:dens_temp_star} (also Figure~\ref{fig:Haro11_cont} for dust), which shows that knot B is overlapping with the cold, dusty disc arm (the 'ear'), while knot C is coinciding with a dense gas clump.

\begin{table}
    \centering
    \caption{Predicted escaping emission of the three stellar knots at present-day (0\,Myr). Calculated as described in Section~\ref{sec:mock_method}. }
    \begin{tabular}{c|c|c|c|c}
        \hline
        & Knot A & Knot B & Knot C \\
        \hline
       $L_{\rm LyC}$ [erg\,s$^{-1}$] & $8\times10^{40}$ &  $3\times10^{41}$ &  $2\times10^{42}$ \\
       $L_{\rm Ly\alpha}$ [erg\,s$^{-1}$] &  $3\times10^{38}$ &  $6\times10^{38}$ &  $6\times10^{39}$ \\
       $L_{\rm H\alpha}$ [erg\,s$^{-1}$] & $2\times10^{37}$ &  $2\times10^{40}$ &  $4\times10^{39}$ \\
       \hline
       $f_{\rm esc}^{\rm LyC}$ [\%]  & 1.1 & 1.7 & 1.4 \\
       $f_{\rm esc}^{\rm Ly\alpha}$ [\%] & 57 & 0.3 & 26 \\   % 127 & 0.29 & 26 \\
       $f_{\rm esc}^{\rm H\alpha}$ [\%] & $\approx100$ & 9.6 & $\approx 100$   %%% & 123 & 9.6 & 137   
       %% Total easc for each emission is 1.5, 6, 33 (as seen in figure 6)
    \end{tabular}
    \label{tab:f_esc}
\end{table}

%%%%%%%%%%%%%%%%%%%%%%%%
%%%%%%%%%%%%%%%%%%%%%%%%
%%%%%%%%%%%%%%%%%%%%%%%%
\subsection{Caveats}\label{sec:caveats}

%%%%%%%%%%%%%%%%%%%%%%%%
% The escape fraction of \Ha{} in our simulation is lower than expected when comparing to the attenuation of star-forming galaxies in the local \citep[e.g.][]{Groves+12} and high-redshift Universe \citep[$z\sim 2-3$; e.g.][]{Reddy+20}. This could be due to our galaxy actively undergoing a merger, as very high attenuation has been found in extreme and dusty environment \citep[e.g.][]{Garcia-Marin+09}. However, there are also a few numerical effects that could boost the luminosity to expected levels, such as a stronger feedback recipe, top-heavy initial mass function, or different dust recipe. 

One of the aims of this study is to reproduce and understand properties of Haro 11, for which we discuss possible caveats here. For example, the \Ha{} escape fraction in our simulation is lower than what can be inferred from observations of attenuation in Haro 11. Observations show that Haro 11 has a highly structured, patchy ISM with substantial nebular extinction concentrated around its knots \citep[e.g.][]{Hayes+07, James+13, Menacho+21}, yet its global attenuation remains moderate compared to the deeply embedded star formation in our simulated merger. This difference may be due to the adopted physical prescriptions, such as feedback strength, IMF assumptions, or the adopted dust model, which can influence both the intrinsic and emergent \Ha{} luminosities.

% Moreover, the low \Ha{} escape fraction suggests that the amount of dust might be overestimated; particularly because the intrinsic \Ha{} luminosity yields the expected SFR. Our simulation is modelled after the metal-poor Haro 11 galaxy, but we have assumed standard treatment of dust. a more tailored application of dust handling could increase both the luminosity and escape fraction of \Ha{} (and \Lya{} \& LyC). For example
% In particular, adopting a different attenuation law, metal-to-dust conversion, or use a non-Solar mix of metals, could decrease the amount of extinction. Moreover, observations indicate that the dust-to-gas mass ratio is a function of metallicity and following a power-law \citep[][]{Remy-Ruyer+14}. \citet[][]{Mauerhofer+21} found that very young stars do not contribute much to the spectra, due to suppression from the dusty star-forming regions; meaning much of their ionising radiation might be absorbed by dust before being able to ionise the ISM. 

In particular, adopting a different attenuation law, metal-to-dust conversion, or non-Solar metal mixture could decrease the amount of extinction, thus increasing the escaping luminosities. Observationally, the dust content is known to correlate with metallicity, with the dust-to-gas mass ratio following a power-law relation \citep[][]{Remy-Ruyer+14}. This has important implications for ionising radiation, as \citet[][]{Mauerhofer+21} showed that very young stars contribute little to the observed spectra because they remain embedded in dusty star-forming regions, where much of their LyC radiation is absorbed before reaching the ISM. In a cosmological zoom-in of three dwarf galaxies, \citet[][]{Trebitsch+17} found that the LyC is primarily absorbed in nearby clouds (within 100 parsec), after which it will escape into the galaxy halo.

%%% As \Ha{} emission is directly connected to feedback physics and dust recipes 
% Highly attenuated, which is highly unusual \citep[ the Balmer decrement in][]{Groves+12} but is has been observed in dusty cores, like in galaxy merger \citep[][]{}. 

%%% The escaping \Ha\ luminosity is a factor of $\lesssim 10$ lower than expected for the SFR of our simulated galaxy, using conventional conversion factors, e.g. ${\rm SFR} = L_{\rm H\alpha} \times 7.9\times10^{-42}\,{\rm M_\odot\,yr^{-1}\,erg^{-1}\,s}$ \citep[][]{Kennicutt98}
%%% This could be explained by the combined numerical uncertainties of dust calculations, assumed SED model and metallicity-dependence \citep[compared to][]{Kennicutt98}, assuming instantaneous star formation, and treating \Ha{} propagation as instantaneous in the post-processing.

%%% SED fits, star formation recipe, stellar feedback routine, and \Ha\ emissivity calculations; as well as inherent variance of the conversion factor

% stronger early feedback could increase the escape fraction by creating escape channels in the natal clouds of bright young stars.

Additionally, the low \Ha{} escape fraction we find suggests that young, bright stars are deeply embedded in their dusty natal clouds. While radiation feedback is included in our simulation, its impact is likely underestimated because the cloud-scale structure and early expansion of \HII{} regions are not fully resolved, limiting the ability of young stars to clear escape channels. For example, cloud-scale simulations by \citet[][]{Kimm+19} suggest that radiation feedback is important in disrupting molecular clouds before supernovae, at $\sim2\,$Myr, and feedback in general is important in disrupting clouds \citep[e.g.][]{Kimm+17, Trebitsch+17}. Furthermore, \citet[][]{Smith+22} discuss the survival of dust in ionised gas in simulated disc-like galaxies, and their results indicate that the presence of dust within ionised gas can have a notable impact on the escape fraction; particularly within \Ha{} bright, \HII{} regions. Recent observations by \citet[][]{Carr+25} suggest that early feedback is critical for enabling LyC escape, as multiphase supernova-driven winds can otherwise obscure the emerging radiation. Similarly, \citet[][]{Komarova+25} find that radiation feedback is often enough to explain the emergent LyC escape in LzLCS galaxies.  % \citep[see also][for LzLCS leakers with lack of evidence of recent supernovae]{Bait+24}.
% possible that early radiative feedback is not strong enough to clear channels,

In particular, if supernovae are the main mechanism that facilitate escape channels, a $\sim10\,$Myr delay would be expected between star formation and a surge in $f_{\rm esc}$ \citep[see e.g.][]{KimmCen2014, Trebitsch+17}. However, early radiative feedback from massive stars can result in a time delay as little as only a few Myr \citep[e.g.][]{Ma+16, Kimm+17, Kimm+19}. Our results show no significant time delay between the peaks in SFR and the increase in escape fraction. This could be caused by efficient early feedback or perhaps due to merger-induced star formation in combination with ram pressure stripping (during the pericentres). However, it is difficult (and beyond the scope of this project) to discern what process is facilitating escape (and the exact time delay) as the time resolution of our outputs are $\Delta t = 5\,$Myr (with the exception around $t\sim0\,$Myr). 
%  for clearing and ionising gas to form 

% When the Strömgren radius is not resolved by enough simulation cells, radiation artificially couples to too much mass.
%%% the feedback energy from younger stars becomes smoothed over a larger volume. 
Our simulation has not included any recipe to treat unresolved Strömgren spheres, which represent the physical extent of H{\small II} regions. When the Strömgren radius is not resolved by enough simulation cells, radiation artificially couples to too much mass. This results in cells with an unphysical mix of warm ($\sim8000\,$K) gas that is partially ionised, a known effect of under-resolving ionisation fronts \citep[e.g.][]{Rosdahl+15, Jaura+20, Deng+24}. Consequently, the lower ionisation fraction reduces recombination emissivities, since recombination scales as $j\propto n_e n_p$. However, \citet[][]{Deng+24} have shown that the partially ionised gas from unresolved Strömgren spheres result in overestimation of the ionised gas mass, which can boost \Ha{} emissivities \citep[see also][]{Ejdetjarn+24}. Additional complex interactions, such as fewer ionised escape channels and increased neutral fraction (and perhaps dust content), further complicate \Lya{} transfer and attenuation. In our case, the intrinsic \Lya{} and \Ha{} luminosities lie close to Haro 11 observations and match values calculated using conventional conversion factors from the SFR. However, phase diagrams of our simulation indicate the presence of unresolved H{\small II} regions that might shift the luminosities and worsen the match.

Simulating specific galaxies is difficult due to their complex and dynamic nature, and is increasingly difficult for detailed structures, which can be highly stochastic. This is especially true for galaxy mergers, which involves a high number of degrees of freedom and the resulting galaxy properties are sensitive to the initial conditions. Therefore, it is encouraging that our simulations are able to reproduce key features of Haro 11 (see also E25), since similar gas and stellar conditions yield similar spectra. Although quantitative comparisons are not feasible at this stage, qualitative comparisons and generalisations allows specific galaxy simulations, like this one, to be applied to a broader understanding of similar, local starburst galaxies. In particular, our simulation offers a detailed view of the evolution of radiation escape during a galaxy merger.

%%%%%%%%%%%%%%%%%%%%%%%%%%%%%%%%%%%%%%%%%%%%%%%%%%
\section{Conclusions}\label{sec:conclusions}
We have performed radiation hydrodynamics simulations of a galaxy merger with key features similar to Haro 11 (based on initial conditions from E25) and performed post-processing to track the propagation of LyC, \Lya, and \Ha{} radiation. Our results highlight the potential impact of mergers on the escape of radiation, insightful as a case study of interacting galaxies and as a qualitative comparison with Haro 11 observations. Our conclusions are as follows:
\begin{itemize}
    \item Galaxy interactions have a noticeable effect on the escape of LyC. During the close interactions of the merger \emph{and} during SFR peaks, there is an order of magnitude increase in the LyC escape fraction. The boost at the pericentres is likely due to a combination of dynamical (interaction) and stellar feedback effects. In particular, we discuss the idea that ram pressure can cause displacement of stars into regions of low gas density. We suggest that dynamical effects play a bigger role during the first interaction, as the increases in SFR and intrinsic LyC are moderately low at this time. The other periods of increased LyC escape are associated with bursts in star formation and, thus, stellar feedback will be more relevant.
    % , through feedback creating escape channels, stripping of neutral gas, and geometry 
    % Notably, while the SFR is higher after the first pericentre, the escape fraction is roughly the same as at that pericentre. 
    \item \Lya{} follows a similar evolution to LyC, but with a more shallow increase at the SFR peaks and low variation across different sightlines. This can be attributed to collisionally excited \Lya{} outside of the galaxy, which are less affected by dust. Conversely, the \Ha{} escape fraction decreases during the SFR peaks. We suggest this can be due to newborn stars being deeply embedded into their (dusty) natal clouds.
    
    \item The LyC luminosity and escape fraction varies by roughly two orders of magnitude purely due to variation in observation angle. Thus, significant LyC leakage from a BCG, such as Haro 11, is highly dependent on a favourable angle. However, during the first pericentre we found that LyC leakage increased along most sightlines, which suggest that dynamical effects can boost the visibility of LyC leakers.
    \item Our simulation captures several features that are observed in Haro 11, including the production of LyC in knot B \& C as well as the larger escape of LyC from knot C and \Lya{} line profile features. In particular, we find a broad agreement of the overall \Lya\ spectral shape in the knots between observations of Haro 11 and predicted by our mocks. The specifics of the \Lya\ line profile are highly sensitive to the conditions of the gas, and a qualitative match implies these conditions are similar to those in Haro 11, which is encouraging for direct comparisons with our simulation. % Any qualitative match beteen observations and a simulation offers credence to its validity :)
\end{itemize}{}

Overall, our simulation demonstrates that galaxy mergers can help facilitate the escape of LyC, through both dynamical effects (e.g. stellar displacement and gas stripping) and localised feedback processes. The strong dependence on the viewing-angle for LyC escape highlights the importance of gas geometry and orientation in LyC emitting galaxies. However, our simulations imply that merger effects might improve the visibility of Lyman continuum emitters from several different sightlines. Our results encourage further numerical studies aimed at disentangling the interplay between dynamics, feedback, and radiative transfer in mergers, in order to better understand how LyC and \Lya{} escape is shaped by galaxy interactions; particularly in the highly interacting systems now being uncovered in the JWST era.

% Our results encourage further numerical studies aimed at disentangling the interplay between dynamics, feedback, and radiative transfer in galaxy mergers, to better understand how LyC and \Lya\ escape is affected by merger and interactions. In particular, to understand the first, highly-interacting, galaxies observed during the JWST era.
%and how to relate this back to observed galaxy properties.

%%%%%%%%%%%%%%%%%%%%%%%%%%%%%%%%%%%%%%%%%%%%%%%%%%
\section*{Acknowledgements}
This project was supported by foundations managed by The Royal Swedish Academy of Sciences. The computations and data handling were enabled by resources provided by the National Academic Infrastructure for Supercomputing in Sweden (NAISS), partially funded by the Swedish Research Council through grant agreement no. 2022-06725. This project made use of the medium computation allocations and small storage allocation of NAISS.

OA acknowledges support from the Knut and Alice Wallenberg Foundation, The Swedish Research Council (grant 2025-04892), the Swedish National Space Agency (SNSA Dnr 2023-00164), the LMK foundation, and eSSENCE, a Swedish strategic research programme in e-Science.

Data analysis made use of the \texttt{yt-project} \citep[][]{Turk+11} to analyse \ramses outputs. As well as \texttt{numpy} \citep[][]{Harris+20} and \texttt{matplotlib} \citep[][]{Hunter+07}, for analysis with \texttt{Python} version 3.9 (Python Software Foundation, https://www.python.org/).

% GÖ/JR/JB/OA acknowledges grant from ...

%%%%%%%%%%%%%%%%%%%%%%%%%%%%%%%%%%%%%%%%%%%%%%%%%%
\section*{Data Availability}
The data underlying this article will be shared on reasonable request to the corresponding author.
 
% The inclusion of a Data Availability Statement is a requirement for articles published in MNRAS. Data Availability Statements provide a standardised format for readers to understand the availability of data underlying the research results described in the article. The statement may refer to original data generated in the course of the study or to third-party data analysed in the article. The statement should describe and provide means of access, where possible, by linking to the data or providing the required accession numbers for the relevant databases or DOIs.



\bibliographystyle{mnras}
%\bibliography{complete.bib} % if your bibtex file is called example.bib

\begin{thebibliography}{}
\makeatletter
\relax
\def\mn@urlcharsother{\let\do\@makeother \do\$\do\&\do\#\do\^\do\_\do\%\do\~}
\def\mn@doi{\begingroup\mn@urlcharsother \@ifnextchar [ {\mn@doi@} {\mn@doi@[]}}
\def\mn@doi@[#1]#2{\def\@tempa{#1}\ifx\@tempa\@empty \href {http://dx.doi.org/#2} {doi:#2}\else \href {http://dx.doi.org/#2} {#1}\fi \endgroup}
\def\mn@eprint#1#2{\mn@eprint@#1:#2::\@nil}
\def\mn@eprint@arXiv#1{\href {http://arxiv.org/abs/#1} {{\tt arXiv:#1}}}
\def\mn@eprint@dblp#1{\href {http://dblp.uni-trier.de/rec/bibtex/#1.xml} {dblp:#1}}
\def\mn@eprint@#1:#2:#3:#4\@nil{\def\@tempa {#1}\def\@tempb {#2}\def\@tempc {#3}\ifx \@tempc \@empty \let \@tempc \@tempb \let \@tempb \@tempa \fi \ifx \@tempb \@empty \def\@tempb {arXiv}\fi \@ifundefined {mn@eprint@\@tempb}{\@tempb:\@tempc}{\expandafter \expandafter \csname mn@eprint@\@tempb\endcsname \expandafter{\@tempc}}}

\bibitem[\protect\citeauthoryear{{Adamo}, {{\"O}stlin}, {Zackrisson}, {Hayes}, {Cumming}  \& {Micheva}}{{Adamo} et~al.}{2010}]{Adamo+10}
{Adamo} A.,  {{\"O}stlin} G.,  {Zackrisson} E.,  {Hayes} M.,  {Cumming} R.~J.,   {Micheva} G.,  2010, \mn@doi [\mnras] {10.1111/j.1365-2966.2010.16983.x}, \href {https://ui.adsabs.harvard.edu/abs/2010MNRAS.407..870A} {407, 870}

\bibitem[\protect\citeauthoryear{{Agertz} \& {Kravtsov}}{{Agertz} \& {Kravtsov}}{2015}]{AgertzKravtsov15}
{Agertz} O.,  {Kravtsov} A.~V.,  2015, \mn@doi [\apj] {10.1088/0004-637X/804/1/18}, \href {http://adsabs.harvard.edu/abs/2015ApJ...804...18A} {804, 18}

\bibitem[\protect\citeauthoryear{{Agertz} \& {Kravtsov}}{{Agertz} \& {Kravtsov}}{2016}]{AgertzKravtsov16}
{Agertz} O.,  {Kravtsov} A.~V.,  2016, \mn@doi [\apj] {10.3847/0004-637X/824/2/79}, \href {http://adsabs.harvard.edu/abs/2016ApJ...824...79A} {824, 79}

\bibitem[\protect\citeauthoryear{{Agertz}, {Kravtsov}, {Leitner}  \& {Gnedin}}{{Agertz} et~al.}{2013}]{Agertz+13}
{Agertz} O.,  {Kravtsov} A.~V.,  {Leitner} S.~N.,   {Gnedin} N.~Y.,  2013, \mn@doi [\apj] {10.1088/0004-637X/770/1/25}, \href {http://adsabs.harvard.edu/abs/2013ApJ...770...25A} {770, 25}

\bibitem[\protect\citeauthoryear{{Agertz} et~al.,}{{Agertz} et~al.}{2020}]{Agertz+20}
{Agertz} O.,  et~al., 2020, \mn@doi [\mnras] {10.1093/mnras/stz3053}, \href {https://ui.adsabs.harvard.edu/abs/2020MNRAS.491.1656A} {491, 1656}

\bibitem[\protect\citeauthoryear{{Agertz} et~al.,}{{Agertz} et~al.}{2021}]{Agertz+21}
{Agertz} O.,  et~al., 2021, \mn@doi [\mnras] {10.1093/mnras/stab322}, \href {https://ui.adsabs.harvard.edu/abs/2021MNRAS.503.5826A} {503, 5826}

\bibitem[\protect\citeauthoryear{{Anderson}, {Governato}, {Karcher}, {Quinn}  \& {Wadsley}}{{Anderson} et~al.}{2017}]{Anderson+17}
{Anderson} L.,  {Governato} F.,  {Karcher} M.,  {Quinn} T.,   {Wadsley} J.,  2017, \mn@doi [\mnras] {10.1093/mnras/stx709}, \href {https://ui.adsabs.harvard.edu/abs/2017MNRAS.468.4077A} {468, 4077}

\bibitem[\protect\citeauthoryear{{Asplund}, {Grevesse}, {Sauval}  \& {Scott}}{{Asplund} et~al.}{2009}]{Asplund2009}
{Asplund} M.,  {Grevesse} N.,  {Sauval} A.~J.,   {Scott} P.,  2009, \mn@doi [\araa] {10.1146/annurev.astro.46.060407.145222}, \href {http://adsabs.harvard.edu/abs/2009ARA\%26A..47..481A} {47, 481}

\bibitem[\protect\citeauthoryear{{Atek} et~al.,}{{Atek} et~al.}{2024}]{Atek+24}
{Atek} H.,  et~al., 2024, \mn@doi [\nat] {10.1038/s41586-024-07043-6}, \href {https://ui.adsabs.harvard.edu/abs/2024Natur.626..975A} {626, 975}

\bibitem[\protect\citeauthoryear{{Bait} et~al.,}{{Bait} et~al.}{2024}]{Omkar+24}
{Bait} O.,  et~al., 2024, \mn@doi [\aap] {10.1051/0004-6361/202348416}, \href {https://ui.adsabs.harvard.edu/abs/2024A&A...688A.198B} {688, A198}

\bibitem[\protect\citeauthoryear{{Bergvall}, {Zackrisson}, {Andersson}, {Arnberg}, {Masegosa}  \& {{\"O}stlin}}{{Bergvall} et~al.}{2006}]{Bergvall+06}
{Bergvall} N.,  {Zackrisson} E.,  {Andersson} B.~G.,  {Arnberg} D.,  {Masegosa} J.,   {{\"O}stlin} G.,  2006, \mn@doi [\aap] {10.1051/0004-6361:20053788}, \href {https://ui.adsabs.harvard.edu/abs/2006A&A...448..513B} {448, 513}

\bibitem[\protect\citeauthoryear{{Bergvall}, {Leitet}, {Zackrisson}  \& {Marquart}}{{Bergvall} et~al.}{2013}]{Bergvall+13}
{Bergvall} N.,  {Leitet} E.,  {Zackrisson} E.,   {Marquart} T.,  2013, \mn@doi [\aap] {10.1051/0004-6361/201118433}, \href {https://ui.adsabs.harvard.edu/abs/2013A&A...554A..38B} {554, A38}

\bibitem[\protect\citeauthoryear{{Bigiel}, {Leroy}, {Walter}, {Brinks}, {de Blok}, {Madore}  \& {Thornley}}{{Bigiel} et~al.}{2008}]{Bigiel+08}
{Bigiel} F.,  {Leroy} A.,  {Walter} F.,  {Brinks} E.,  {de Blok} W.~J.~G.,  {Madore} B.,   {Thornley} M.~D.,  2008, \mn@doi [\aj] {10.1088/0004-6256/136/6/2846}, \href {http://adsabs.harvard.edu/abs/2008AJ....136.2846B} {136, 2846}

\bibitem[\protect\citeauthoryear{{Blaizot} et~al.,}{{Blaizot} et~al.}{2023}]{Blaizot+23}
{Blaizot} J.,  et~al., 2023, \mn@doi [\mnras] {10.1093/mnras/stad1523}, \href {https://ui.adsabs.harvard.edu/abs/2023MNRAS.523.3749B} {523, 3749}

\bibitem[\protect\citeauthoryear{{Bosman} et~al.,}{{Bosman} et~al.}{2022}]{Bosman+22}
{Bosman} S. E.~I.,  et~al., 2022, \mn@doi [\mnras] {10.1093/mnras/stac1046}, \href {https://ui.adsabs.harvard.edu/abs/2022MNRAS.514...55B} {514, 55}

\bibitem[\protect\citeauthoryear{{Bruzual} \& {Charlot}}{{Bruzual} \& {Charlot}}{2003}]{BruzualCharlot03}
{Bruzual} G.,  {Charlot} S.,  2003, \mn@doi [\mnras] {10.1046/j.1365-8711.2003.06897.x}, \href {https://ui.adsabs.harvard.edu/abs/2003MNRAS.344.1000B} {344, 1000}

\bibitem[\protect\citeauthoryear{{Carr} et~al.,}{{Carr} et~al.}{2025}]{Carr+25}
{Carr} C.,  et~al., 2025, \mn@doi [arXiv e-prints] {10.48550/arXiv.2510.21197}, \href {https://ui.adsabs.harvard.edu/abs/2025arXiv251021197C} {p. arXiv:2510.21197}

\bibitem[\protect\citeauthoryear{{Chabrier}}{{Chabrier}}{2003}]{Chabrier03}
{Chabrier} G.,  2003, \mn@doi [Publications of the Astronomical Society of the Pacific] {10.1086/376392}, \href {https://ui.adsabs.harvard.edu/\#abs/2003PASP..115..763C} {115, 763}

\bibitem[\protect\citeauthoryear{{Choustikov} et~al.,}{{Choustikov} et~al.}{2024}]{Choustikov+24}
{Choustikov} N.,  et~al., 2024, \mn@doi [\mnras] {10.1093/mnras/stae1586}, \href {https://ui.adsabs.harvard.edu/abs/2024MNRAS.532.2463C} {532, 2463}

\bibitem[\protect\citeauthoryear{{Citro} et~al.,}{{Citro} et~al.}{2025}]{Citro+25}
{Citro} A.,  et~al., 2025, \mn@doi [\apj] {10.3847/1538-4357/add5e6}, \href {https://ui.adsabs.harvard.edu/abs/2025ApJ...986..184C} {986, 184}

\bibitem[\protect\citeauthoryear{{Deharveng}, {Buat}, {Le Brun}, {Milliard}, {Kunth}, {Shull}  \& {Gry}}{{Deharveng} et~al.}{2001}]{Deharveng+01}
{Deharveng} J.~M.,  {Buat} V.,  {Le Brun} V.,  {Milliard} B.,  {Kunth} D.,  {Shull} J.~M.,   {Gry} C.,  2001, \mn@doi [\aap] {10.1051/0004-6361:20010920}, \href {https://ui.adsabs.harvard.edu/abs/2001A&A...375..805D} {375, 805}

\bibitem[\protect\citeauthoryear{{Deng}, {Li}, {Kannan}, {Smith}, {Vogelsberger}  \& {Bryan}}{{Deng} et~al.}{2024}]{Deng+24}
{Deng} Y.,  {Li} H.,  {Kannan} R.,  {Smith} A.,  {Vogelsberger} M.,   {Bryan} G.~L.,  2024, \mn@doi [\mnras] {10.1093/mnras/stad3202}, \href {https://ui.adsabs.harvard.edu/abs/2024MNRAS.527..478D} {527, 478}

\bibitem[\protect\citeauthoryear{{Dijkstra}}{{Dijkstra}}{2017}]{Dijkstra17}
{Dijkstra} M.,  2017, arXiv e-prints, \href {https://ui.adsabs.harvard.edu/abs/2017arXiv170403416D} {p. arXiv:1704.03416}

\bibitem[\protect\citeauthoryear{{Duncan} et~al.,}{{Duncan} et~al.}{2019}]{Duncan+19}
{Duncan} K.,  et~al., 2019, \mn@doi [\apj] {10.3847/1538-4357/ab148a}, \href {https://ui.adsabs.harvard.edu/abs/2019ApJ...876..110D} {876, 110}

\bibitem[\protect\citeauthoryear{{Ejdetj{\"a}rn}, {Agertz}, {{\"O}stlin}, {Rey}  \& {Renaud}}{{Ejdetj{\"a}rn} et~al.}{2024}]{Ejdetjarn+24}
{Ejdetj{\"a}rn} T.,  {Agertz} O.,  {{\"O}stlin} G.,  {Rey} M.~P.,   {Renaud} F.,  2024, \mn@doi [\mnras] {10.1093/mnras/stae2099}, \href {https://ui.adsabs.harvard.edu/abs/2024MNRAS.534..135E} {534, 135}

\bibitem[\protect\citeauthoryear{{Ejdetj{\"a}rn}, {Agertz}, {Renaud}, {{\"O}stlin}, {Le Reste}  \& {Adamo}}{{Ejdetj{\"a}rn} et~al.}{2025}]{Ejdetjarn+25}
{Ejdetj{\"a}rn} T.,  {Agertz} O.,  {Renaud} F.,  {{\"O}stlin} G.,  {Le Reste} A.,   {Adamo} A.,  2025, \mn@doi [\mnras] {10.1093/mnras/staf1733}, \href {https://ui.adsabs.harvard.edu/abs/2025MNRAS.543.3849E} {543, 3849}

\bibitem[\protect\citeauthoryear{{Emsellem} \& {van de Ven}}{{Emsellem} \& {van de Ven}}{2008}]{Emsellem+08}
{Emsellem} E.,  {van de Ven} G.,  2008, \mn@doi [\apj] {10.1086/524720}, \href {https://ui.adsabs.harvard.edu/abs/2008ApJ...674..653E} {674, 653}

\bibitem[\protect\citeauthoryear{{Ferland}, {Korista}, {Verner}, {Ferguson}, {Kingdon}  \& {Verner}}{{Ferland} et~al.}{1998}]{Ferland+98}
{Ferland} G.~J.,  {Korista} K.~T.,  {Verner} D.~A.,  {Ferguson} J.~W.,  {Kingdon} J.~B.,   {Verner} E.~M.,  1998, \mn@doi [\pasp] {10.1086/316190}, \href {https://ui.adsabs.harvard.edu/abs/1998PASP..110..761F} {110, 761}

\bibitem[\protect\citeauthoryear{{Finkelstein} et~al.,}{{Finkelstein} et~al.}{2019}]{Finkelstein+19}
{Finkelstein} S.~L.,  et~al., 2019, \mn@doi [\apj] {10.3847/1538-4357/ab1ea8}, \href {https://ui.adsabs.harvard.edu/abs/2019ApJ...879...36F} {879, 36}

\bibitem[\protect\citeauthoryear{{Flury} et~al.,}{{Flury} et~al.}{2022a}]{Flury+22a}
{Flury} S.~R.,  et~al., 2022a, \mn@doi [\apjs] {10.3847/1538-4365/ac5331}, \href {https://ui.adsabs.harvard.edu/abs/2022ApJS..260....1F} {260, 1}

\bibitem[\protect\citeauthoryear{{Flury} et~al.,}{{Flury} et~al.}{2022b}]{Flury+22b}
{Flury} S.~R.,  et~al., 2022b, \mn@doi [\apj] {10.3847/1538-4357/ac61e4}, \href {https://ui.adsabs.harvard.edu/abs/2022ApJ...930..126F} {930, 126}

\bibitem[\protect\citeauthoryear{{Flury} et~al.,}{{Flury} et~al.}{2024}]{Flury+24}
{Flury} S.~R.,  et~al., 2024, \mn@doi [arXiv e-prints] {10.48550/arXiv.2409.12118}, \href {https://ui.adsabs.harvard.edu/abs/2024arXiv240912118F} {p. arXiv:2409.12118}

\bibitem[\protect\citeauthoryear{{Gao} et~al.,}{{Gao} et~al.}{2022}]{Gao+22}
{Gao} Y.,  et~al., 2022, \mn@doi [\aap] {10.1051/0004-6361/202142309}, \href {https://ui.adsabs.harvard.edu/abs/2022A&A...661A.136G} {661, A136}

\bibitem[\protect\citeauthoryear{{Grimes} et~al.,}{{Grimes} et~al.}{2007}]{Grimes+07}
{Grimes} J.~P.,  et~al., 2007, \mn@doi [\apj] {10.1086/521353}, \href {https://ui.adsabs.harvard.edu/abs/2007ApJ...668..891G} {668, 891}

\bibitem[\protect\citeauthoryear{{Grisdale}, {Agertz}, {Renaud}  \& {Romeo}}{{Grisdale} et~al.}{2018}]{Grisdale2018}
{Grisdale} K.,  {Agertz} O.,  {Renaud} F.,   {Romeo} A.~B.,  2018, \mn@doi [\mnras] {10.1093/mnras/sty1595}, \href {http://adsabs.harvard.edu/abs/2018MNRAS.tmp.1523G} {}

\bibitem[\protect\citeauthoryear{{Grisdale}, {Agertz}, {Renaud}, {Romeo}, {Devriendt}  \& {Slyz}}{{Grisdale} et~al.}{2019}]{Grisdale2019}
{Grisdale} K.,  {Agertz} O.,  {Renaud} F.,  {Romeo} A.~B.,  {Devriendt} J.,   {Slyz} A.,  2019, arXiv e-prints, \href {http://adsabs.harvard.edu/abs/2019arXiv190200518G} {}

\bibitem[\protect\citeauthoryear{Harris et~al.,}{Harris et~al.}{2020}]{Harris+20}
Harris C.~R.,  et~al., 2020, \mn@doi [Nature] {10.1038/s41586-020-2649-2}, 585, 357–362

\bibitem[\protect\citeauthoryear{{Hayes}, {{\"O}stlin}, {Atek}, {Kunth}, {Mas-Hesse}, {Leitherer}, {Jim{\'e}nez-Bail{\'o}n}  \& {Adamo}}{{Hayes} et~al.}{2007}]{Hayes+07}
{Hayes} M.,  {{\"O}stlin} G.,  {Atek} H.,  {Kunth} D.,  {Mas-Hesse} J.~M.,  {Leitherer} C.,  {Jim{\'e}nez-Bail{\'o}n} E.,   {Adamo} A.,  2007, \mn@doi [\mnras] {10.1111/j.1365-2966.2007.12482.x}, \href {https://ui.adsabs.harvard.edu/abs/2007MNRAS.382.1465H} {382, 1465}

\bibitem[\protect\citeauthoryear{{Hernquist}}{{Hernquist}}{1990}]{Hernquist1990}
{Hernquist} L.,  1990, \mn@doi [\apj] {10.1086/168845}, \href {http://adsabs.harvard.edu/abs/1990ApJ...356..359H} {356, 359}

\bibitem[\protect\citeauthoryear{Hunter}{Hunter}{2007}]{Hunter+07}
Hunter J.~D.,  2007, \mn@doi [Computing in Science \& Engineering] {10.1109/MCSE.2007.55}, 9, 90

\bibitem[\protect\citeauthoryear{{Inami} et~al.,}{{Inami} et~al.}{2022}]{Inami+22}
{Inami} H.,  et~al., 2022, \mn@doi [\mnras] {10.1093/mnras/stac1779}, \href {https://ui.adsabs.harvard.edu/abs/2022MNRAS.515.3126I} {515, 3126}

\bibitem[\protect\citeauthoryear{{Inoue} \& {Iwata}}{{Inoue} \& {Iwata}}{2008}]{InoueIwata2008}
{Inoue} A.~K.,  {Iwata} I.,  2008, \mn@doi [\mnras] {10.1111/j.1365-2966.2008.13350.x}, \href {https://ui.adsabs.harvard.edu/abs/2008MNRAS.387.1681I} {387, 1681}

\bibitem[\protect\citeauthoryear{{Inoue}, {Shimizu}, {Iwata}  \& {Tanaka}}{{Inoue} et~al.}{2014}]{Inoue+14}
{Inoue} A.~K.,  {Shimizu} I.,  {Iwata} I.,   {Tanaka} M.,  2014, \mn@doi [\mnras] {10.1093/mnras/stu936}, \href {https://ui.adsabs.harvard.edu/abs/2014MNRAS.442.1805I} {442, 1805}

\bibitem[\protect\citeauthoryear{{Izotov}, {Worseck}, {Schaerer}, {Guseva}, {Thuan}, {Fricke}  \& {Orlitov{\'a}}}{{Izotov} et~al.}{2018}]{Izotov+18}
{Izotov} Y.~I.,  {Worseck} G.,  {Schaerer} D.,  {Guseva} N.~G.,  {Thuan} T.~X.,  {Fricke} A. V.,   {Orlitov{\'a}} I.,  2018, \mn@doi [\mnras] {10.1093/mnras/sty1378}, \href {https://ui.adsabs.harvard.edu/abs/2018MNRAS.478.4851I} {478, 4851}

\bibitem[\protect\citeauthoryear{{Izotov}, {Worseck}, {Schaerer}, {Guseva}, {Chisholm}, {Thuan}, {Fricke}  \& {Verhamme}}{{Izotov} et~al.}{2021}]{Izotov+21}
{Izotov} Y.~I.,  {Worseck} G.,  {Schaerer} D.,  {Guseva} N.~G.,  {Chisholm} J.,  {Thuan} T.~X.,  {Fricke} K.~J.,   {Verhamme} A.,  2021, \mn@doi [\mnras] {10.1093/mnras/stab612}, \href {https://ui.adsabs.harvard.edu/abs/2021MNRAS.503.1734I} {503, 1734}

\bibitem[\protect\citeauthoryear{{James}, {Tsamis}, {Walsh}, {Barlow}  \& {Westmoquette}}{{James} et~al.}{2013}]{James+13}
{James} B.~L.,  {Tsamis} Y.~G.,  {Walsh} J.~R.,  {Barlow} M.~J.,   {Westmoquette} M.~S.,  2013, \mn@doi [\mnras] {10.1093/mnras/stt034}, \href {https://ui.adsabs.harvard.edu/abs/2013MNRAS.430.2097J} {430, 2097}

\bibitem[\protect\citeauthoryear{{Jaura}, {Magg}, {Glover}  \& {Klessen}}{{Jaura} et~al.}{2020}]{Jaura+20}
{Jaura} O.,  {Magg} M.,  {Glover} S. C.~O.,   {Klessen} R.~S.,  2020, \mn@doi [\mnras] {10.1093/mnras/staa3054}, \href {https://ui.adsabs.harvard.edu/abs/2020MNRAS.499.3594J} {499, 3594}

\bibitem[\protect\citeauthoryear{{Kennicutt}}{{Kennicutt}}{1998}]{Kennicutt98}
{Kennicutt} Jr. R.~C.,  1998, \mn@doi [\araa] {10.1146/annurev.astro.36.1.189}, \href {http://adsabs.harvard.edu/abs/1998ARA\%26A..36..189K} {36, 189}

\bibitem[\protect\citeauthoryear{{Kim} \& {Ostriker}}{{Kim} \& {Ostriker}}{2015}]{KimOstriker15}
{Kim} C.-G.,  {Ostriker} E.~C.,  2015, \mn@doi [\apj] {10.1088/0004-637X/802/2/99}, \href {https://ui.adsabs.harvard.edu/\#abs/2015ApJ...802...99K} {802, 99}

\bibitem[\protect\citeauthoryear{{Kimm} \& {Cen}}{{Kimm} \& {Cen}}{2014}]{KimmCen2014}
{Kimm} T.,  {Cen} R.,  2014, \mn@doi [\apj] {10.1088/0004-637X/788/2/121}, \href {https://ui.adsabs.harvard.edu/abs/2014ApJ...788..121K} {788, 121}

\bibitem[\protect\citeauthoryear{{Kimm}, {Katz}, {Haehnelt}, {Rosdahl}, {Devriendt}  \& {Slyz}}{{Kimm} et~al.}{2017}]{Kimm+17}
{Kimm} T.,  {Katz} H.,  {Haehnelt} M.,  {Rosdahl} J.,  {Devriendt} J.,   {Slyz} A.,  2017, \mn@doi [\mnras] {10.1093/mnras/stx052}, \href {https://ui.adsabs.harvard.edu/abs/2017MNRAS.466.4826K} {466, 4826}

\bibitem[\protect\citeauthoryear{{Kimm}, {Blaizot}, {Garel}, {Michel-Dansac}, {Katz}, {Rosdahl}, {Verhamme}  \& {Haehnelt}}{{Kimm} et~al.}{2019}]{Kimm+19}
{Kimm} T.,  {Blaizot} J.,  {Garel} T.,  {Michel-Dansac} L.,  {Katz} H.,  {Rosdahl} J.,  {Verhamme} A.,   {Haehnelt} M.,  2019, \mn@doi [\mnras] {10.1093/mnras/stz989}, \href {https://ui.adsabs.harvard.edu/abs/2019MNRAS.486.2215K} {486, 2215}

\bibitem[\protect\citeauthoryear{{Komarova} et~al.,}{{Komarova} et~al.}{2024}]{Komarova+24}
{Komarova} L.,  et~al., 2024, \mn@doi [\apj] {10.3847/1538-4357/ad3962}, \href {https://ui.adsabs.harvard.edu/abs/2024ApJ...967..117K} {967, 117}

\bibitem[\protect\citeauthoryear{{Komarova} et~al.,}{{Komarova} et~al.}{2025}]{Komarova+25}
{Komarova} L.,  et~al., 2025, \mn@doi [\apj] {10.3847/1538-4357/ae0e0a}, \href {https://ui.adsabs.harvard.edu/abs/2025ApJ...994..192K} {994, 192}

\bibitem[\protect\citeauthoryear{{Kostyuk} \& {Ciardi}}{{Kostyuk} \& {Ciardi}}{2024}]{KostyukCiardi2024}
{Kostyuk} I.,  {Ciardi} B.,  2024, arXiv e-prints, \href {https://ui.adsabs.harvard.edu/abs/2024arXiv241204348K} {p. arXiv:2412.04348}

\bibitem[\protect\citeauthoryear{{Krumholz} \& {Tan}}{{Krumholz} \& {Tan}}{2007}]{KrumholzTan07}
{Krumholz} M.~R.,  {Tan} J.~C.,  2007, \mn@doi [\apj] {10.1086/509101}, \href {https://ui.adsabs.harvard.edu/\#abs/2007ApJ...654..304K} {654, 304}

\bibitem[\protect\citeauthoryear{{Laursen}, {Sommer-Larsen}  \& {Andersen}}{{Laursen} et~al.}{2009}]{Laursen+09}
{Laursen} P.,  {Sommer-Larsen} J.,   {Andersen} A.~C.,  2009, \mn@doi [\apj] {10.1088/0004-637X/704/2/1640}, \href {https://ui.adsabs.harvard.edu/abs/2009ApJ...704.1640L} {704, 1640}

\bibitem[\protect\citeauthoryear{{Le Reste} et~al.,}{{Le Reste} et~al.}{2024}]{LeReste+24}
{Le Reste} A.,  et~al., 2024, \mn@doi [\mnras] {10.1093/mnras/stad3910}, \href {https://ui.adsabs.harvard.edu/abs/2024MNRAS.528..757L} {528, 757}

\bibitem[\protect\citeauthoryear{{Le Reste} et~al.,}{{Le Reste} et~al.}{2025a}]{LeReste+25c}
{Le Reste} A.,  et~al., 2025a, \mn@doi [arXiv e-prints] {10.48550/arXiv.2509.06922}, \href {https://ui.adsabs.harvard.edu/abs/2025arXiv250906922L} {p. arXiv:2509.06922}

\bibitem[\protect\citeauthoryear{{Le Reste} et~al.,}{{Le Reste} et~al.}{2025b}]{LeReste+25b}
{Le Reste} A.,  et~al., 2025b, \mn@doi [\apjs] {10.3847/1538-4365/adf227}, \href {https://ui.adsabs.harvard.edu/abs/2025ApJS..280...27L} {280, 27}

\bibitem[\protect\citeauthoryear{{Leclercq} et~al.,}{{Leclercq} et~al.}{2024}]{Leclercq+24}
{Leclercq} F.,  et~al., 2024, \mn@doi [\aap] {10.1051/0004-6361/202449362}, \href {https://ui.adsabs.harvard.edu/abs/2024A&A...687A..73L} {687, A73}

\bibitem[\protect\citeauthoryear{{Leitet}, {Bergvall}, {Piskunov}  \& {Andersson}}{{Leitet} et~al.}{2011}]{Leitet+11}
{Leitet} E.,  {Bergvall} N.,  {Piskunov} N.,   {Andersson} B.~G.,  2011, \mn@doi [\aap] {10.1051/0004-6361/201015654}, \href {https://ui.adsabs.harvard.edu/abs/2011A&A...532A.107L} {532, A107}

\bibitem[\protect\citeauthoryear{{Leitet}, {Bergvall}, {Hayes}, {Linn{\'e}}  \& {Zackrisson}}{{Leitet} et~al.}{2013}]{Leitet+13}
{Leitet} E.,  {Bergvall} N.,  {Hayes} M.,  {Linn{\'e}} S.,   {Zackrisson} E.,  2013, \mn@doi [\aap] {10.1051/0004-6361/201118370}, \href {https://ui.adsabs.harvard.edu/abs/2013A&A...553A.106L} {553, A106}

\bibitem[\protect\citeauthoryear{{Leitherer}, {Hernandez}, {Lee}  \& {Oey}}{{Leitherer} et~al.}{2016}]{Leitherer+16}
{Leitherer} C.,  {Hernandez} S.,  {Lee} J.~C.,   {Oey} M.~S.,  2016, \mn@doi [\apj] {10.3847/0004-637X/823/1/64}, \href {https://ui.adsabs.harvard.edu/abs/2016ApJ...823...64L} {823, 64}

\bibitem[\protect\citeauthoryear{{Levermore}}{{Levermore}}{1984}]{Levermore84}
{Levermore} C.~D.,  1984, \mn@doi [\jqsrt] {10.1016/0022-4073(84)90112-2}, \href {https://ui.adsabs.harvard.edu/abs/1984JQSRT..31..149L} {31, 149}

\bibitem[\protect\citeauthoryear{{Lewis} et~al.,}{{Lewis} et~al.}{2020}]{Lewis+20}
{Lewis} J. S.~W.,  et~al., 2020, \mn@doi [\mnras] {10.1093/mnras/staa1748}, \href {https://ui.adsabs.harvard.edu/abs/2020MNRAS.496.4342L} {496, 4342}

\bibitem[\protect\citeauthoryear{{Li} \& {Draine}}{{Li} \& {Draine}}{2001}]{LiDraine2001}
{Li} A.,  {Draine} B.~T.,  2001, \mn@doi [\apj] {10.1086/323147}, \href {https://ui.adsabs.harvard.edu/abs/2001ApJ...554..778L} {554, 778}

\bibitem[\protect\citeauthoryear{{Ma}, {Hopkins}, {Kasen}, {Quataert}, {Faucher-Gigu{\`e}re}, {Kere{\v{s}}}, {Murray}  \& {Strom}}{{Ma} et~al.}{2016}]{Ma+16}
{Ma} X.,  {Hopkins} P.~F.,  {Kasen} D.,  {Quataert} E.,  {Faucher-Gigu{\`e}re} C.-A.,  {Kere{\v{s}}} D.,  {Murray} N.,   {Strom} A.,  2016, \mn@doi [\mnras] {10.1093/mnras/stw941}, \href {https://ui.adsabs.harvard.edu/abs/2016MNRAS.459.3614M} {459, 3614}

\bibitem[\protect\citeauthoryear{{MacHattie}, {Irwin}, {Madden}, {Cormier}  \& {R{\'e}my-Ruyer}}{{MacHattie} et~al.}{2014}]{MacHattie+14}
{MacHattie} J.~A.,  {Irwin} J.~A.,  {Madden} S.~C.,  {Cormier} D.,   {R{\'e}my-Ruyer} A.,  2014, \mn@doi [\mnras] {10.1093/mnrasl/slt160}, \href {https://ui.adsabs.harvard.edu/abs/2014MNRAS.438L..66M} {438, L66}

\bibitem[\protect\citeauthoryear{{Madau}}{{Madau}}{1995}]{Madau1995}
{Madau} P.,  1995, \mn@doi [\apj] {10.1086/175332}, \href {https://ui.adsabs.harvard.edu/abs/1995ApJ...441...18M} {441, 18}

\bibitem[\protect\citeauthoryear{{Madden} et~al.,}{{Madden} et~al.}{2013}]{Madden+13}
{Madden} S.~C.,  et~al., 2013, \mn@doi [\pasp] {10.1086/671138}, \href {https://ui.adsabs.harvard.edu/abs/2013PASP..125..600M} {125, 600}

\bibitem[\protect\citeauthoryear{{Maji} et~al.,}{{Maji} et~al.}{2022}]{Maji+22}
{Maji} M.,  et~al., 2022, \mn@doi [\aap] {10.1051/0004-6361/202142740}, \href {https://ui.adsabs.harvard.edu/abs/2022A&A...663A..66M} {663, A66}

\bibitem[\protect\citeauthoryear{{Mascia} et~al.,}{{Mascia} et~al.}{2025}]{Mascia+25}
{Mascia} S.,  et~al., 2025, \mn@doi [arXiv e-prints] {10.48550/arXiv.2501.08268}, \href {https://ui.adsabs.harvard.edu/abs/2025arXiv250108268M} {p. arXiv:2501.08268}

\bibitem[\protect\citeauthoryear{{Matsuoka} et~al.,}{{Matsuoka} et~al.}{2018}]{Matsuoka+18}
{Matsuoka} Y.,  et~al., 2018, \mn@doi [\apj] {10.3847/1538-4357/aaee7a}, \href {https://ui.adsabs.harvard.edu/abs/2018ApJ...869..150M} {869, 150}

\bibitem[\protect\citeauthoryear{{Matthee} et~al.,}{{Matthee} et~al.}{2022}]{Matthee+22}
{Matthee} J.,  et~al., 2022, \mn@doi [\mnras] {10.1093/mnras/stac801}, \href {https://ui.adsabs.harvard.edu/abs/2022MNRAS.512.5960M} {512, 5960}

\bibitem[\protect\citeauthoryear{{Mauerhofer}, {Verhamme}, {Blaizot}, {Garel}, {Kimm}, {Michel-Dansac}  \& {Rosdahl}}{{Mauerhofer} et~al.}{2021}]{Mauerhofer+21}
{Mauerhofer} V.,  {Verhamme} A.,  {Blaizot} J.,  {Garel} T.,  {Kimm} T.,  {Michel-Dansac} L.,   {Rosdahl} J.,  2021, \mn@doi [\aap] {10.1051/0004-6361/202039449}, \href {https://ui.adsabs.harvard.edu/abs/2021A&A...646A..80M} {646, A80}

\bibitem[\protect\citeauthoryear{{McGreer}, {Mesinger}  \& {D'Odorico}}{{McGreer} et~al.}{2015}]{McGreer+15}
{McGreer} I.~D.,  {Mesinger} A.,   {D'Odorico} V.,  2015, \mn@doi [\mnras] {10.1093/mnras/stu2449}, \href {https://ui.adsabs.harvard.edu/abs/2015MNRAS.447..499M} {447, 499}

\bibitem[\protect\citeauthoryear{{Menacho} et~al.,}{{Menacho} et~al.}{2021}]{Menacho+21}
{Menacho} V.,  et~al., 2021, \mn@doi [\mnras] {10.1093/mnras/stab1491}, \href {https://ui.adsabs.harvard.edu/abs/2021MNRAS.506.1777M} {506, 1777}

\bibitem[\protect\citeauthoryear{{Michel-Dansac}, {Blaizot}, {Garel}, {Verhamme}, {Kimm}  \& {Trebitsch}}{{Michel-Dansac} et~al.}{2020}]{Michel-Dansac+20}
{Michel-Dansac} L.,  {Blaizot} J.,  {Garel} T.,  {Verhamme} A.,  {Kimm} T.,   {Trebitsch} M.,  2020, \mn@doi [\aap] {10.1051/0004-6361/201834961}, \href {https://ui.adsabs.harvard.edu/abs/2020A&A...635A.154M} {635, A154}

\bibitem[\protect\citeauthoryear{{Naidu} et~al.,}{{Naidu} et~al.}{2022}]{Naidu+22}
{Naidu} R.~P.,  et~al., 2022, \mn@doi [\mnras] {10.1093/mnras/stab3601}, \href {https://ui.adsabs.harvard.edu/abs/2022MNRAS.510.4582N} {510, 4582}

\bibitem[\protect\citeauthoryear{{Navarro}, {Frenk}  \& {White}}{{Navarro} et~al.}{1996}]{NavarroFrenkWhite96}
{Navarro} J.~F.,  {Frenk} C.~S.,   {White} S.~D.~M.,  1996, \mn@doi [\apj] {10.1086/177173}, \href {http://adsabs.harvard.edu/abs/1996ApJ...462..563N} {462, 563}

\bibitem[\protect\citeauthoryear{{{\"O}stlin}, {Amram}, {Bergvall}, {Masegosa}, {Boulesteix}  \& {M{\'a}rquez}}{{{\"O}stlin} et~al.}{2001}]{Ostlin+01}
{{\"O}stlin} G.,  {Amram} P.,  {Bergvall} N.,  {Masegosa} J.,  {Boulesteix} J.,   {M{\'a}rquez} I.,  2001, \mn@doi [\aap] {10.1051/0004-6361:20010832}, \href {https://ui.adsabs.harvard.edu/abs/2001A&A...374..800O} {374, 800}

\bibitem[\protect\citeauthoryear{{{\"O}stlin}, {Marquart}, {Cumming}, {Fathi}, {Bergvall}, {Adamo}, {Amram}  \& {Hayes}}{{{\"O}stlin} et~al.}{2015}]{Ostlin+15}
{{\"O}stlin} G.,  {Marquart} T.,  {Cumming} R.~J.,  {Fathi} K.,  {Bergvall} N.,  {Adamo} A.,  {Amram} P.,   {Hayes} M.,  2015, \mn@doi [\aap] {10.1051/0004-6361/201323233}, \href {https://ui.adsabs.harvard.edu/abs/2015A&A...583A..55O} {583, A55}

\bibitem[\protect\citeauthoryear{{{\"O}stlin} et~al.,}{{{\"O}stlin} et~al.}{2021}]{Ostlin+21}
{{\"O}stlin} G.,  et~al., 2021, \mn@doi [\apj] {10.3847/1538-4357/abf1e8}, \href {https://ui.adsabs.harvard.edu/abs/2021ApJ...912..155O} {912, 155}

\bibitem[\protect\citeauthoryear{{R{\'e}my-Ruyer} et~al.,}{{R{\'e}my-Ruyer} et~al.}{2014}]{Remy-Ruyer+14}
{R{\'e}my-Ruyer} A.,  et~al., 2014, \mn@doi [\aap] {10.1051/0004-6361/201322803}, \href {https://ui.adsabs.harvard.edu/abs/2014A&A...563A..31R} {563, A31}

\bibitem[\protect\citeauthoryear{{Renaud}, {Agertz}, {Andersson}, {Read}, {Ryde}, {Bensby}, {Rey}  \& {Feuillet}}{{Renaud} et~al.}{2021}]{Renaud+21a}
{Renaud} F.,  {Agertz} O.,  {Andersson} E.~P.,  {Read} J.~I.,  {Ryde} N.,  {Bensby} T.,  {Rey} M.~P.,   {Feuillet} D.~K.,  2021, \mn@doi [\mnras] {10.1093/mnras/stab543}, \href {https://ui.adsabs.harvard.edu/abs/2021MNRAS.503.5868R} {503, 5868}

\bibitem[\protect\citeauthoryear{{Rivera-Thorsen} et~al.,}{{Rivera-Thorsen} et~al.}{2017a}]{Rivera-Thorsen+17b}
{Rivera-Thorsen} T.~E.,  et~al., 2017a, \mn@doi [\aap] {10.1051/0004-6361/201732173}, \href {https://ui.adsabs.harvard.edu/abs/2017A&A...608L...4R} {608, L4}

\bibitem[\protect\citeauthoryear{{Rivera-Thorsen}, {{\"O}stlin}, {Hayes}  \& {Puschnig}}{{Rivera-Thorsen} et~al.}{2017b}]{Rivera-Thorsen+17a}
{Rivera-Thorsen} T.~E.,  {{\"O}stlin} G.,  {Hayes} M.,   {Puschnig} J.,  2017b, \mn@doi [\apj] {10.3847/1538-4357/aa5d0a}, \href {https://ui.adsabs.harvard.edu/abs/2017ApJ...837...29R} {837, 29}

\bibitem[\protect\citeauthoryear{{Robertson}, {Ellis}, {Furlanetto}  \& {Dunlop}}{{Robertson} et~al.}{2015}]{Robertson+15}
{Robertson} B.~E.,  {Ellis} R.~S.,  {Furlanetto} S.~R.,   {Dunlop} J.~S.,  2015, \mn@doi [\apjl] {10.1088/2041-8205/802/2/L19}, \href {https://ui.adsabs.harvard.edu/abs/2015ApJ...802L..19R} {802, L19}

\bibitem[\protect\citeauthoryear{{Rodriguez-Gomez} et~al.,}{{Rodriguez-Gomez} et~al.}{2015}]{Rodriguez-Gomez+15}
{Rodriguez-Gomez} V.,  et~al., 2015, \mn@doi [\mnras] {10.1093/mnras/stv264}, \href {https://ui.adsabs.harvard.edu/abs/2015MNRAS.449...49R} {449, 49}

\bibitem[\protect\citeauthoryear{{Rosdahl} \& {Teyssier}}{{Rosdahl} \& {Teyssier}}{2015}]{RosdahlTeyssier15}
{Rosdahl} J.,  {Teyssier} R.,  2015, \mn@doi [\mnras] {10.1093/mnras/stv567}, \href {https://ui.adsabs.harvard.edu/abs/2015MNRAS.449.4380R} {449, 4380}

\bibitem[\protect\citeauthoryear{{Rosdahl}, {Blaizot}, {Aubert}, {Stranex}  \& {Teyssier}}{{Rosdahl} et~al.}{2013}]{Rosdahl+13}
{Rosdahl} J.,  {Blaizot} J.,  {Aubert} D.,  {Stranex} T.,   {Teyssier} R.,  2013, \mn@doi [\mnras] {10.1093/mnras/stt1722}, \href {https://ui.adsabs.harvard.edu/abs/2013MNRAS.436.2188R} {436, 2188}

\bibitem[\protect\citeauthoryear{Rosdahl, Schaye, Teyssier  \& Agertz}{Rosdahl et~al.}{2015}]{Rosdahl+15}
Rosdahl J.,  Schaye J.,  Teyssier R.,   Agertz O.,  2015, \mn@doi [Monthly Notices of the Royal Astronomical Society] {10.1093/mnras/stv937}, 451, 34

\bibitem[\protect\citeauthoryear{{Rosdahl} et~al.,}{{Rosdahl} et~al.}{2022}]{Rosdahl+22}
{Rosdahl} J.,  et~al., 2022, \mn@doi [\mnras] {10.1093/mnras/stac1942}, \href {https://ui.adsabs.harvard.edu/abs/2022MNRAS.515.2386R} {515, 2386}

\bibitem[\protect\citeauthoryear{{Rosen} \& {Bregman}}{{Rosen} \& {Bregman}}{1995}]{RosenBregman95}
{Rosen} A.,  {Bregman} J.~N.,  1995, \mn@doi [\apj] {10.1086/175303}, \href {http://adsabs.harvard.edu/abs/1995ApJ...440..634R} {440, 634}

\bibitem[\protect\citeauthoryear{{Saldana-Lopez} et~al.,}{{Saldana-Lopez} et~al.}{2022}]{Saldana-Lopez+22}
{Saldana-Lopez} A.,  et~al., 2022, \mn@doi [\aap] {10.1051/0004-6361/202141864}, \href {https://ui.adsabs.harvard.edu/abs/2022A&A...663A..59S} {663, A59}

\bibitem[\protect\citeauthoryear{{Saldana-Lopez} et~al.,}{{Saldana-Lopez} et~al.}{2023}]{Saldana-Lopez+23}
{Saldana-Lopez} A.,  et~al., 2023, \mn@doi [\mnras] {10.1093/mnras/stad1283}, \href {https://ui.adsabs.harvard.edu/abs/2023MNRAS.522.6295S} {522, 6295}

\bibitem[\protect\citeauthoryear{{Saldana-Lopez} et~al.,}{{Saldana-Lopez} et~al.}{2025}]{Saldana-Lopez+25}
{Saldana-Lopez} A.,  et~al., 2025, \mn@doi [arXiv e-prints] {10.48550/arXiv.2504.07074}, \href {https://ui.adsabs.harvard.edu/abs/2025arXiv250407074S} {p. arXiv:2504.07074}

\bibitem[\protect\citeauthoryear{{Schaerer}, {Izotov}, {Verhamme}, {Orlitov{\'a}}, {Thuan}, {Worseck}  \& {Guseva}}{{Schaerer} et~al.}{2016}]{Schaerer+16}
{Schaerer} D.,  {Izotov} Y.~I.,  {Verhamme} A.,  {Orlitov{\'a}} I.,  {Thuan} T.~X.,  {Worseck} G.,   {Guseva} N.~G.,  2016, \mn@doi [\aap] {10.1051/0004-6361/201628943}, \href {https://ui.adsabs.harvard.edu/abs/2016A&A...591L...8S} {591, L8}

\bibitem[\protect\citeauthoryear{{Schmidt}}{{Schmidt}}{1959}]{Schmidt59}
{Schmidt} M.,  1959, \mn@doi [\apj] {10.1086/146614}, \href {http://adsabs.harvard.edu/abs/1959ApJ...129..243S} {129, 243}

\bibitem[\protect\citeauthoryear{{Simmonds} et~al.,}{{Simmonds} et~al.}{2024}]{Simmonds+24}
{Simmonds} C.,  et~al., 2024, \mn@doi [\mnras] {10.1093/mnras/stad3605}, \href {https://ui.adsabs.harvard.edu/abs/2024MNRAS.527.6139S} {527, 6139}

\bibitem[\protect\citeauthoryear{{Sirressi} et~al.,}{{Sirressi} et~al.}{2022}]{Sirressi+22}
{Sirressi} M.,  et~al., 2022, \mn@doi [\mnras] {10.1093/mnras/stab3774}, \href {https://ui.adsabs.harvard.edu/abs/2022MNRAS.510.4819S} {510, 4819}

\bibitem[\protect\citeauthoryear{{Smith} et~al.,}{{Smith} et~al.}{2022}]{Smith+22}
{Smith} A.,  et~al., 2022, \mn@doi [\mnras] {10.1093/mnras/stac2641}, \href {https://ui.adsabs.harvard.edu/abs/2022MNRAS.517....1S} {517, 1}

\bibitem[\protect\citeauthoryear{{Solhaug} et~al.,}{{Solhaug} et~al.}{2025}]{Solhaug+25}
{Solhaug} E.,  et~al., 2025, \mn@doi [The Open Journal of Astrophysics] {10.33232/001c.134065}, \href {https://ui.adsabs.harvard.edu/abs/2025OJAp....8E..35S} {8, 35}

\bibitem[\protect\citeauthoryear{{Steidel}, {Bogosavljevi{\'c}}, {Shapley}, {Reddy}, {Rudie}, {Pettini}, {Trainor}  \& {Strom}}{{Steidel} et~al.}{2018}]{Steidel+18}
{Steidel} C.~C.,  {Bogosavljevi{\'c}} M.,  {Shapley} A.~E.,  {Reddy} N.~A.,  {Rudie} G.~C.,  {Pettini} M.,  {Trainor} R.~F.,   {Strom} A.~L.,  2018, \mn@doi [\apj] {10.3847/1538-4357/aaed28}, \href {https://ui.adsabs.harvard.edu/abs/2018ApJ...869..123S} {869, 123}

\bibitem[\protect\citeauthoryear{{Teyssier}}{{Teyssier}}{2002}]{Teyssier2002}
{Teyssier} R.,  2002, \mn@doi [\aap] {10.1051/0004-6361:20011817}, \href {https://ui.adsabs.harvard.edu/abs/2002A&A...385..337T} {385, 337}

\bibitem[\protect\citeauthoryear{{Toro}, {Spruce}  \& {Speares}}{{Toro} et~al.}{1994}]{ToroSpruceSpeares94}
{Toro} E.~F.,  {Spruce} M.,   {Speares} W.,  1994, \mn@doi [Shock Waves] {10.1007/BF01414629}, \href {https://ui.adsabs.harvard.edu/abs/1994ShWav...4...25T} {4, 25}

\bibitem[\protect\citeauthoryear{{Trebitsch}, {Blaizot}, {Rosdahl}, {Devriendt}  \& {Slyz}}{{Trebitsch} et~al.}{2017}]{Trebitsch+17}
{Trebitsch} M.,  {Blaizot} J.,  {Rosdahl} J.,  {Devriendt} J.,   {Slyz} A.,  2017, \mn@doi [\mnras] {10.1093/mnras/stx1060}, \href {https://ui.adsabs.harvard.edu/abs/2017MNRAS.470..224T} {470, 224}

\bibitem[\protect\citeauthoryear{{Turk}, {Smith}, {Oishi}, {Skory}, {Skillman}, {Abel}  \& {Norman}}{{Turk} et~al.}{2011}]{Turk+11}
{Turk} M.~J.,  {Smith} B.~D.,  {Oishi} J.~S.,  {Skory} S.,  {Skillman} S.~W.,  {Abel} T.,   {Norman} M.~L.,  2011, \mn@doi [\apjs] {10.1088/0067-0049/192/1/9}, \href {http://adsabs.harvard.edu/abs/2011ApJS..192....9T} {192, 9}

\bibitem[\protect\citeauthoryear{{Vanzella} et~al.,}{{Vanzella} et~al.}{2018}]{Vanzella+18}
{Vanzella} E.,  et~al., 2018, \mn@doi [\mnras] {10.1093/mnrasl/sly023}, \href {https://ui.adsabs.harvard.edu/abs/2018MNRAS.476L..15V} {476, L15}

\bibitem[\protect\citeauthoryear{{Verhamme}, {Orlitov{\'a}}, {Schaerer}  \& {Hayes}}{{Verhamme} et~al.}{2015}]{Verhamme+15}
{Verhamme} A.,  {Orlitov{\'a}} I.,  {Schaerer} D.,   {Hayes} M.,  2015, \mn@doi [\aap] {10.1051/0004-6361/201423978}, \href {https://ui.adsabs.harvard.edu/abs/2015A&A...578A...7V} {578, A7}

\bibitem[\protect\citeauthoryear{{Verhamme}, {Orlitov{\'a}}, {Schaerer}, {Izotov}, {Worseck}, {Thuan}  \& {Guseva}}{{Verhamme} et~al.}{2017}]{Verhamme+17}
{Verhamme} A.,  {Orlitov{\'a}} I.,  {Schaerer} D.,  {Izotov} Y.,  {Worseck} G.,  {Thuan} T.~X.,   {Guseva} N.,  2017, \mn@doi [\aap] {10.1051/0004-6361/201629264}, \href {https://ui.adsabs.harvard.edu/abs/2017A&A...597A..13V} {597, A13}

\bibitem[\protect\citeauthoryear{{Wang} et~al.,}{{Wang} et~al.}{2021}]{Wang+21}
{Wang} B.,  et~al., 2021, \mn@doi [\apj] {10.3847/1538-4357/ac0434}, \href {https://ui.adsabs.harvard.edu/abs/2021ApJ...916....3W} {916, 3}

\bibitem[\protect\citeauthoryear{{Zhu}, {Yuan}, {Jiang}, {Zheng}  \& {Lin}}{{Zhu} et~al.}{2024}]{Zhu+24}
{Zhu} S.,  {Yuan} F.-T.,  {Jiang} C.,  {Zheng} Z.-Y.,   {Lin} R.,  2024, \mn@doi [\apjl] {10.3847/2041-8213/ad7b18}, \href {https://ui.adsabs.harvard.edu/abs/2024ApJ...974L..20Z} {974, L20}

\makeatother
\end{thebibliography}
\input{complete.bbl} % if your bibtex file is called example.bib

%\begin{thebibliography}{99}
%\end{thebibliography}

%%%%%%%%%%%%%%%%%%%%%%%%%%%%%%%%%%%%%%%%%%%%%%%%%%

%%%%%%%%%%%%%%%%% APPENDICES %%%%%%%%%%%%%%%%%%%%%

%\appendix

%\section{Appendix}

%%%%%%%%%%%%%%%%%%%%%%%%%%%%%%%%%%%%%%%%%%%%%%%%%%

% Don't change these lines
\bsp	% typesetting comment
\label{lastpage}
\end{document}